\documentclass[manuscript,screen]{acmart}
\AtBeginDocument{%
  \providecommand\BibTeX{{%
    Bib\TeX}}}

\setcopyright{acmlicensed}
\copyrightyear{2018}
\acmYear{2018}
\acmDOI{XXXXXXX.XXXXXXX}
\acmConference[Conference acronym 'XX]{Make sure to enter the correct
  conference title from your rights confirmation email}{June 03--05,
  2018}{Woodstock, NY}
\acmISBN{978-1-4503-XXXX-X/2018/06}

\usepackage{amsmath,amsfonts}
\usepackage{algorithmic}
\usepackage{array}
\usepackage[caption=false,font=normalsize,labelfont=sf,textfont=sf]{subfig}
\usepackage{textcomp}
\usepackage{stfloats}
\usepackage{url}
\usepackage{verbatim}
\usepackage{graphicx}
\usepackage{booktabs}
\def\BibTeX{{\rm B\kern-.05em{\sc i\kern-.025em b}\kern-.08em
    T\kern-.1667em\lower.7ex\hbox{E}\kern-.125emX}}
\usepackage{balance}

\usepackage{hyperref}
\usepackage{xspace}
\usepackage{paralist}

\usepackage{amsfonts}
\usepackage{dsfont}
\usepackage{multirow}
\usepackage{multicol}

\usepackage{siunitx}
\DeclareSIUnit{\pp}{\textit{pp}}
\usepackage{pdflscape}
\usepackage{graphicx}
\usepackage[table]{xcolor}
\usepackage{soul}
\usepackage{xcolor}
\definecolor{codegreen}{rgb}{0,0.6,0}
\definecolor{codegray}{rgb}{0.5,0.5,0.5}
\definecolor{codepurple}{rgb}{0.58,0,0.82}
\definecolor{backcolour}{rgb}{0.95,0.95,0.92}

\newcommand{\method}{\textsc{IntTestGen}\xspace}
\newcommand{\evo}{\textrm{EvoSuite}\xspace}
\newcommand{\panta}{\textrm{PANTA}\xspace}
\newcommand{\dfpj}{\textrm{Deps4J}\xspace}

\newcommand{\dfj}{Defects4J\xspace}

\begin{document}

\title{LLM-based Low-Level Integration Test Generation for Java}


\author{Qinghua Xu}
\email{qinghua.xu@ul.ie}
\orcid{0000-0001-8104-1645}
\affiliation{
\institution{Lero, the Research Ireland Centre for Software, University of Limerick}
\city{Limerick}
\country{Ireland}
}

\author{Guancheng Wang}
\email{guancheng.wang@ul.ie}
\orcid{0000-0002-4338-8813}
\affiliation{
\institution{Lero, the Research Ireland Centre for Software, University of Limerick}
\city{Limerick}
\country{Ireland}
}

\author{Lionel Briand}
\email{lionel.briand@lero.ie}
\orcid{0000-0002-1393-1010}
\affiliation{
\institution{Lero, the Research Ireland Centre for Software, University of Limerick}
\city{Limerick}
\country{Ireland}
}
\affiliation{
  \institution{University of Ottawa}
  \city{Ottawa}
  \country{Canada}
}

\author{Zhaoqiang Guo}
\email{gzq@smail.nju.edu.cn}
\orcid{0000-0001-8971-5755}
\affiliation{
\institution{the State Key Laboratory of Blockchain and Data Security, Zhejiang University}
\city{Hangzhou}
\country{China}
}

\author{Kui Liu}
\email{brucekuiliu@gmail.com}
\orcid{0000-0003-0145-615X}
\affiliation{
\institution{Software Engineering Application Technology Lab
 Huawei}
\city{Shanghai}
\country{China}
}
\renewcommand{\shortauthors}{Xu et al.}

\begin{abstract}
Large language models (LLMs) have shown strong potential for automated test generation, yet most approaches target unit-level testing with mocked dependencies. In contrast, low-level integration testing exercises a class with its real, in-project dependencies, revealing faults that mock-based unit tests miss, such as incorrect object construction, invalid API call sequences, and behavioral mismatches between components. However, generating such tests is substantially harder because it requires correctly instantiating real dependencies and adhering to project-specific constraints, both of which are prone to LLM hallucinations.   We identify two root causes behind these hallucinations: \emph{not knowing}, where the LLM lacks sufficient project-specific context, and \emph{not following}, where the LLM fails to comply with constraints even when they are provided in the prompt.

We present \method, an LLM-based test generation approach that targets low-level integration testing by exercising real project dependencies rather than mocking them. \method is built around two strategies: \textit{context-enriched generation} and \textit{constraint-enforced fixing}. To mitigate \emph{not knowing}, context-enriched generation mines real usage patterns from existing project code and leverages them to generate tests. To mitigate \emph{not following}, constraint-enforced fixing performs two-stage repair under symbol-, protocol-, and iteration-level constraints, using a ClassIndex, a Markov typestate model, and experience memory, respectively.

We evaluate \method against the state-of-the-art baselines (\evo and \panta) on \dfj{} and \dfpj{}, a new post-cutoff benchmark containing more complex recent Java repositories. Experimental results show that IntTestGen improves line coverage by 19.99 and 22.69 percentage points and branch coverage by 24.90 and 15.78 percentage points on the two benchmarks, respectively, and improves mutation score by 13.67 and 0.17 percentage points. Beyond the class under test, \method also exercises more real dependency code, covering 378 and 55 additional lines in dependency classes, respectively. The improvement in test quality comes with higher total token and time costs than the baseline, since \method can continue making progress over more iterations rather than plateauing early. Nevertheless, the cost remains practical, averaging 108.97 seconds and 26.59k tokens per method on \dfj{}, and 69.85 seconds and 25.46k tokens per method on \dfpj{}. Ablation results further confirm that all major components contribute positively to the final performance.
\end{abstract}


\received{20 February 2007}
\received[revised]{12 March 2009}
\received[accepted]{5 June 2009}

\maketitle

\section{Introduction}
\label{sec:intro}
Software testing is essential for identifying defects early in the development lifecycle ~\cite{Beck2022TestExample, candor,sweabs}. While unit testing has long been the cornerstone of automated test generation research, it addresses only part of the testing problem. A unit test exercises a single class or method in isolation, typically using mocking frameworks (e.g., Mockito~\cite{mockito}, PowerMock~\cite{powermock}) to replace real dependencies with simulated ones~\cite{intention,mockornot}. This enables fast, deterministic checks, but it creates a fundamental blind spot: faults arising from \textit{interactions} between components, such as invalid object construction, incorrect API call sequences, and behavior mismatches when a dependency's implementation changes. Consider a streaming XML writer: a unit test with mocked dependencies passes even if the writer calls \texttt{writeStartObjection()} without first invoking \texttt{setNextName()}, because the mock silently accepts any call order. A real dependency, however, throws an \texttt{IllegalStateException}, exposing the protocol violation. Such integration faults are common in practice and motivate integration testing.  
 
In this work, we focus on \textbf{low-level integration testing}: testing a class together with its real, in-project dependencies (i.e., classes within the same codebase that the class under test (CUT) calls or instantiates). Unlike end-to-end integration tests that span multiple services or external systems, low-level integration tests operate within a single module or repository. They complement unit tests by validating that components work together correctly, while remaining fast enough to run in continuous integration (CI) pipelines. Despite their importance, existing automated test generation tools largely ignore low-level integration tests. Traditional automated unit-test generators (e.g., \evo~\cite{Fraser2011EvoSuite:Software} and Randoop~\cite{Pacheco2007Randoop:Java}) can execute real dependencies while testing a target class, but do not explicitly optimize dependency interactions or coverage of dependency code, that is, code the target class depends on.  Recent LLM-based approaches~\cite{panta,candor,mutgen,llm4test} have significantly advanced automated software testing in reducing compilation/execution errors and improving test coverage.  However, they either operate at the method level without handling dependencies ~\cite{candor,mutgen}, or rely on mocking frameworks to isolate the CUT~\cite{panta,intention}. System-level integration testing tools (e.g., EvoMaster~\cite{evomaster}, RESTler~\cite{restler}) target REST API endpoints from API specifications. These tools differ fundamentally from this work, where we generate tests directly from source code to exercise the CUT together with its real, in-project dependencies at the class level. To the best of our knowledge, \textbf{no existing automated approach targets class-level integration test generation}. Generating such tests poses unique challenges: the generator must correctly instantiate dependency objects (including abstract types), resolve imports, and respect stateful API protocols, which requires project-specific knowledge that LLMs do not inherently possess.

\textbf{Not knowing vs. Not following.} We characterize these challenges through two root causes of hallucination. \textit{Not knowing} arises when an LLM lacks project-specific context to correctly construct and use dependencies. In those cases, LLMs tend to fabricate nonexistent factory methods or use incorrect constructor signatures. \textit{Not following} occurs when the model receives correct contextual information but fails to adhere to it, producing tests that violate project-specific constraints such as API call ordering or symbol correctness. These two failure modes demand complementary strategies: enriching generation with mined project knowledge, and enforcing constraints explicitly during repair. 

In this work, we propose \method, an LLM-based approach for generating low-level integration tests. \method follows an iterative \textit{plan--generate--validate--fix} workflow, inspired by recent advances in LLM-based test generation~\cite{candor}. To enable effective low-level integration testing, we introduce two novel strategies that strengthen the generation and repair phases by addressing the limitations identified above.

\begin{compactitem}
    \item \textit{Context-enriched Generation.} We incorporate class usage patterns retrieved by a program slicer into the test generation process. For every dependency of the class under test, the slicer mines real instantiation/usage patterns from the project and injects them into the generator's prompt, so that plausible-but-wrong object instantiations are less frequently generated in the first place.
    \item \textit{Constraint-enforced Fixing}. We introduce a two-stage repair mechanism that first generates a fix from execution feedback and then re-generates under explicit symbol-, protocol-, and iteration-level constraints when violations are identified. In the second stage, the LLM must produce a detailed justification of how the revised fix satisfies all constraints and resolves the failure, enabling stronger constraint adherence during repair.
\end{compactitem}

We evaluate \method against two state-of-the-art baselines representing the main paradigms of automated Java test generation: PANTA~\cite{panta}, a recent LLM-based approach that handles classes with dependencies and follows an iterative generate--validate--fix workflow, and EvoSuite~\cite{Fraser2011EvoSuite:Software,evosuite2014}, a widely used state-of-the-art search-based test generator. We conduct the evaluation on Defects4J, a widely used benchmark for Java test generation~\cite{panta,Hossain2024TOGLL:LLMs} that was also used to evaluate PANTA, enabling a direct comparison with prior results.  To avoid potential data leakage and to evaluate our approach on more realistic multi-module projects, we constructed a new dataset, ~\dfpj, by collecting recent Java projects released after the release of the LLM we employed (Qwen3-Coder). Unlike Defects4J, which consists of single-module projects, \dfpj contains more complex multi-module projects. Since mockless tests are intended to exercise real project dependencies, we also measure project dependency line coverage (DepLC), defined as the number of lines in dependency classes other than the CUT that are covered. Experimental results show that IntTestGen outperforms both baselines on Defects4J: it raises line coverage from 68.83\% (PANTA) and 61.52\% (EvoSuite) to 88.82\%, mutation score from 38.33\% and 35.59\% to 52.00\%, and DepLC from 819 and 699 to 1,197 lines per project. Similar gains hold on Deps4J. This improvement costs less than two minutes per method on average, and an ablation study confirms that all major components contribute positively.

The contributions of this work are as follows:
\begin{compactitem}
    \item We identify \textit{dependency usage patterns} as critical contextual information for mockless unit test generation, and propose a novel context-enriched prompting strategy that augments the LLM with real dependency instantiation and usage examples mined from the target project. This significantly reduces hallucinations caused by insufficient contextual knowledge.
    
    \item We propose a novel \textit{constraint-enforced fixing} mechanism that corrects invalid tests at the symbol, protocol, and iteration levels through a two-stage repair process, in which the second stage requires the model to generate a structured justification explaining how the revised fix resolves the original failure and satisfies all constraints. This mechanism goes beyond test generation and directly targets “not following” hallucinations, a common and key bottleneck in many LLM-based tasks that require reliable adherence to explicit constraints.
    
    \item To the best of our knowledge, \method is the first automated approach to explicitly target low-level integration test generation for Java. Unlike prior work that relies on mocks, is limited to method-level testing, or operates at the system API level, \method systematically addresses the challenges of dependency-aware test generation at the source code level. 
    
    \item We conduct a comprehensive empirical evaluation of both \dfj and \dfpj using line coverage, branch coverage, mutation score, and a new metric, \textit{DepLC}, which measures the amount of dependency code and captures the extent to which dependencies are exercised. Results show that \method consistently outperforms both the LLM-based and the search-based SOTA baselines on the effectiveness metrics (line coverage, branch coverage, and mutation score) while covering substantially more dependency code. 
\end{compactitem}
\section{Motivating Example}
\label{sec:example}
\begin{figure}[t]
  \centering
  \subfloat[Class under test: \texttt{ToXmlGenerator}.\label{fig:motex-cut}]{%
    \includegraphics[width=0.8\columnwidth]{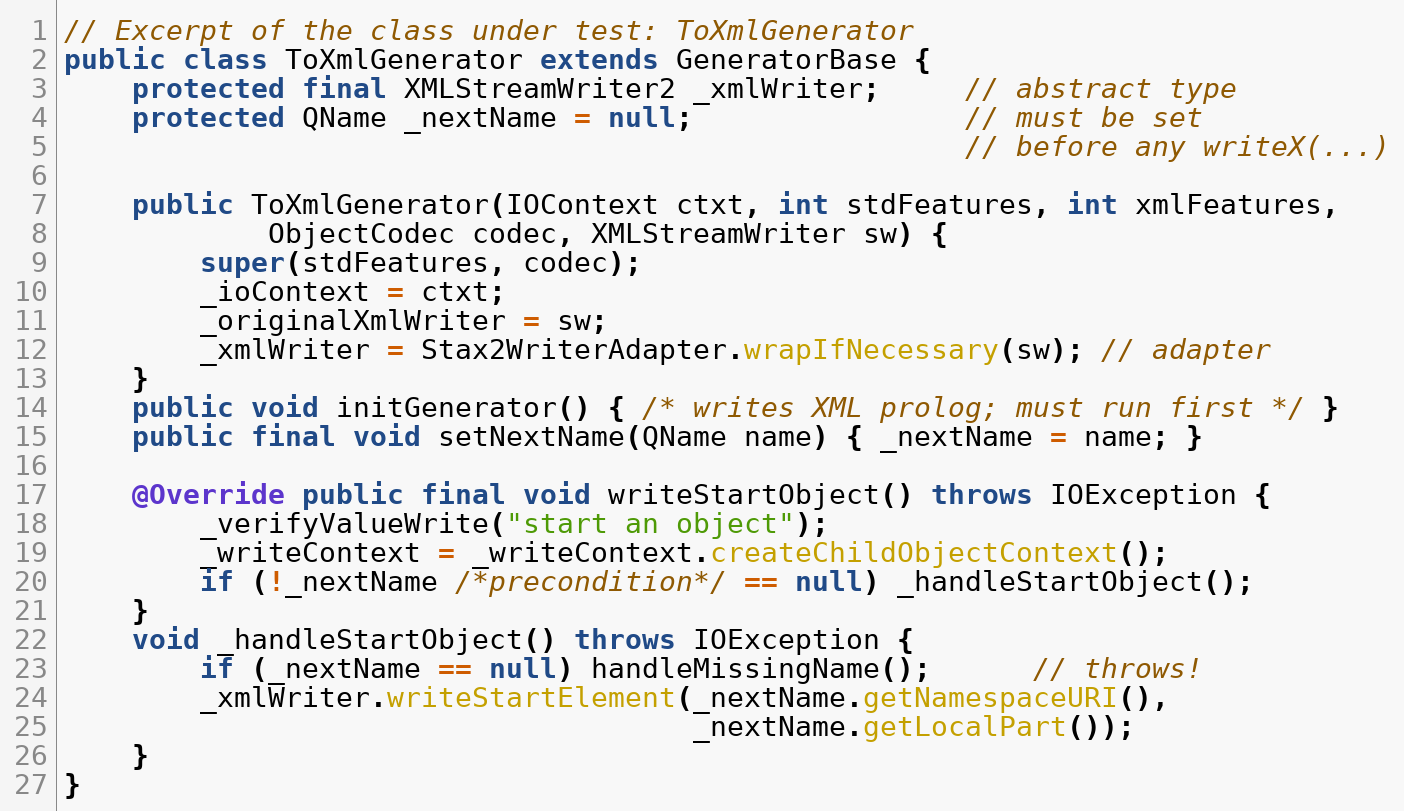}%
  }\\[4pt]
  \subfloat[A plausible but hallucinated LLM attempt
  \label{fig:motex-naive}]{%
    \includegraphics[width=0.8\columnwidth]{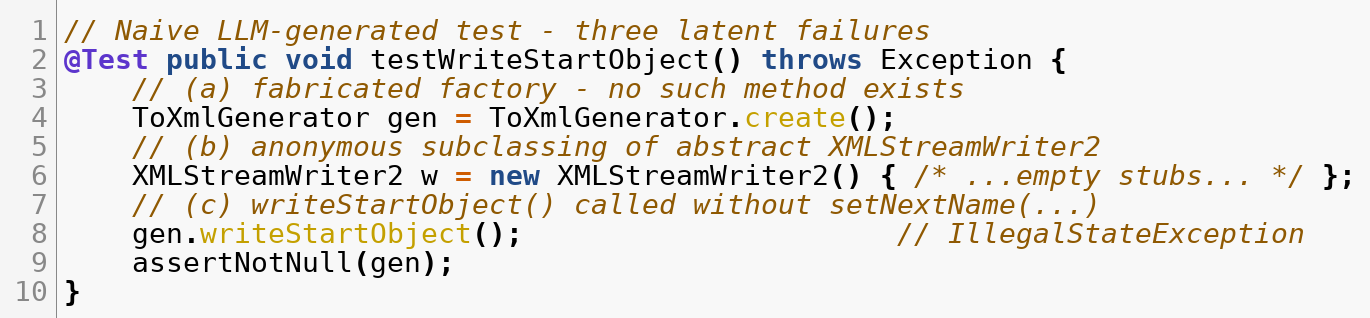}%
  }
  \caption{Motivating example adapted from \dfj.}
  \label{fig:motex}
\end{figure}

We present a motivating example adapted from \dfj in Figure~\ref{fig:motex-cut} and Figure~\ref{fig:motex-naive}. Given the CUT \texttt{ToXmlGenerator}, a valid test must instantiate three external objects: an \texttt{IOContext}, a low-level \texttt{XMLStreamWriter} obtained from \texttt{XMLOutputFactory}, and an internal \texttt{Stax2WriterAdapter} wrapper. In addition, \texttt{setNextName(...)} must be invoked before any \texttt{writeStartX(...)} call to avoid a runtime \texttt{IllegalStateException}. While a unit test using mocking frameworks can bypass these dependency instantiation and usage requirements by simulating external behaviors, it fails to exercise the real interactions among dependent components, which is precisely the gap that integration testing addresses. In contrast, an integration test generator must produce executable tests that instantiate real dependencies and respect valid object usage constraints, making the process substantially more challenging and more prone to invalid test generation. As shown in Figure~\ref{fig:motex-naive}, when an LLM baseline is prompted with only the CUT source code, it generates plausible-looking tests exhibiting two types of generation errors mentioned in Section ~\ref{sec:intro}.

\noindent\textbf{Source~I: Insufficient Context (\textit{``Not Knowing''}).}
The LLM fabricates a \texttt{ToXmlGenerator.create()} factory and emits an empty anonymous subclass for \texttt{XMLStreamWriter2} because it has never seen how real Jackson-XML code constructs the generator or how the abstract writer field is populated. The actual pattern should be allocating an \texttt{XMLStreamWriter} via \texttt{XMLOutputFactory}, passing it to the five-argument constructor, where \texttt{Stax2WriterAdapter.wrapIfNecessary(...)} yields the concrete \texttt{XMLStreamWriter2}. This pattern appears elsewhere in the source code but is absent from the prompt. Such errors arise from \textit{insufficient project-specific knowledge} and thus can be prevented by supplying that knowledge at the input. \method therefore addresses them during generation by mining concrete usage patterns to instantiate an object. 

\noindent\textbf{Source~II: Unenforced Constraints (\textit{``Not Following''}).}
The call \texttt{gen.writeStartObject()} in Figure~\ref{fig:motex-naive} is syntactically correct but fails at runtime because \texttt{\_nextName} was never set before invocation. This reflects a broader failure mode in which the model does not reliably follow project-specific correctness constraints, even when the required information is available. Such failures manifest at three levels. At the \textit{symbol level}, the model may invent non-existent classes or invoke methods with invalid signatures. At the \textit{protocol level}, it may violate required API call sequences, such as omitting \texttt{setNextName(...)} before \texttt{writeStartObject()}. At the \textit{iteration level}, the model may repeatedly generate ineffective fixes because it lacks a mechanism to retain and reuse successful repair patterns across repair attempts. To address this, \method performs a two-stage repair process. The model first generates a fix from the error signals, then checks it against symbol-, protocol-, and iteration-level constraints. If it identifies violations, it triggers a second repair, requiring the model to provide a structured justification explaining why the revised fix satisfies these constraints and resolves the original failure.

\section{Approach}
\label{sec:approach}

\method consists of two stages: a one-time \textit{preparation} stage that constructs project-specific artifacts, and an iterative multi-agent \textit{plan--generate--validate--fix} loop that generates and repairs tests. The iterative loop of \method is designed around two core strategies that address the two sources of limitation identified in Section~\ref{sec:example}. 

In the preparation stage, \method analyzes the target repository to construct three project-specific artifacts that support different parts of the integration testing pipeline. The code property graph (CPG) captures dependency usage in the repository and supports slicing-based retrieval of real object construction and API-call examples. The \textit{ClassIndex} records project-visible classes, methods, constructors, and imports, providing a basis for detecting hallucinated or inconsistent symbols. The initial \textit{Markov typestate model} captures likely valid method-call orders for stateful APIs, providing a basis for detecting illegal call sequences. Together, these artifacts connect context acquisition and constraint enforcement: the CPG helps the model understand how dependencies are used, while the ClassIndex and typestate model help ensure that generated or repaired tests adhere to project-specific symbol and protocol constraints.

It is worth clarifying the scope of project access in \method. The CPG, ClassIndex, and typestate model provide only \emph{construction context}: they inform the Generator and Fixer about valid ways to instantiate and invoke dependencies in the generated test code. At execution time, generated tests exercise the real dependency implementations within the same Maven module as the CUT. \method never instruments, modifies, or mocks dependency behavior; project-wide analysis serves solely to make generated tests compile and execute against real dependencies. This mirrors how a developer writes an integration test: reading related source code to learn the API, then running the test against the actual components.

After preparation, \method enters an iterative \textit{plan--generate--validate--fix} loop composed of five agents: \textsc{Initializer}, \textsc{Planner}, \textsc{Generator}, \textsc{Validator}, and \textsc{Fixer}. The \textsc{Initializer} prepares the analysis artifacts and testing environment,  the \textsc{Planner} creates a test plan by selecting uncovered CUT paths to target, the \textsc{Generator} produces candidate tests, the \textsc{Validator} compiles and executes generated tests, and the \textsc{Fixer} repairs failing tests based on validation feedback. Through this iterative collaboration, the framework progressively improves test quality and coverage over multiple rounds.

Two budgets control the iterative process: $N_{\mathrm{fix}}$ limits the number of repair attempts for each failed test, and $N_{\mathrm{iter}}$ limits the number of full \textit{plan--generate--validate--fix} iterations. The process terminates when the target coverage is reached, when no improvement is observed in two consecutive iterations, or when $N_{\mathrm{iter}}$ is exhausted. The final output is the set of all tests that pass during the iterative loop.

\subsection{Stage I: Preparation}
\label{sec:prelim}

\begin{figure*}[hbt!]
  \centering
  \includegraphics[width=\textwidth]{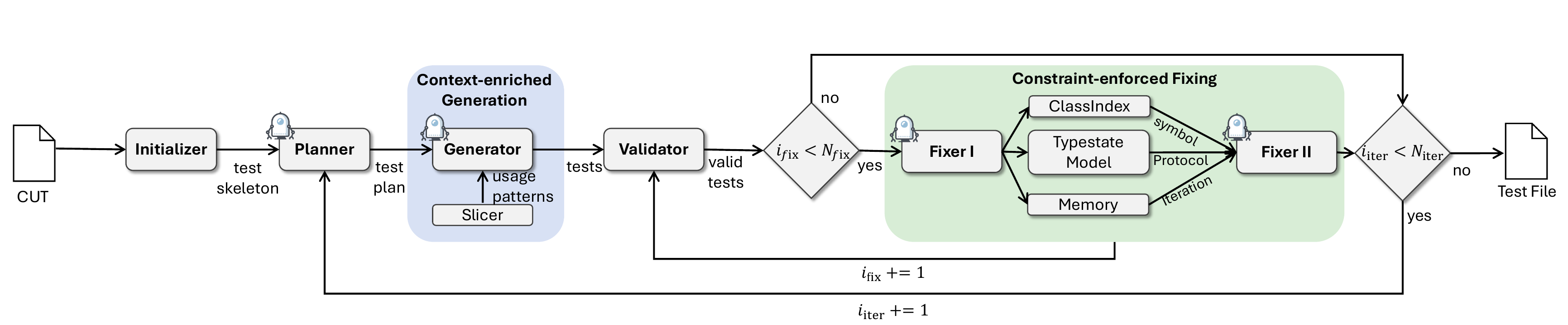}
  \caption{Overview of \method.}
  \label{fig:overview}
\end{figure*}
In this stage, \method constructs the project-specific artifacts needed for the iterative test-generation loop. Specifically, it builds a code property graph (CPG), a \textit{ClassIndex}, and an initial \textit{Markov typestate model}. The CPG and ClassIndex are constructed once and reused throughout the workflow, while the typestate model is initialized here and updated dynamically during subsequent iterations.

\subsubsection{Joern CPG} Joern is a static analysis framework that represents source code as a queryable code property graph (CPG), integrating the abstract syntax tree, intraprocedural control flow, and call/data-flow relations~\cite{joern}. We use Joern because \method requires concrete examples of how project classes are instantiated and invoked in real code, not just type signatures. Specifically, we run Joern on the target repository and use the resulting CPG as the substrate for program slicing, enabling backward slicing to reconstruct real usage patterns of the CUT and its dependencies.

\subsubsection{ClassIndex} The \textit{ClassIndex} records all classes visible under the project's build configuration, together with the members needed to validate generated tests. For each class reachable from the project source tree, test source tree, Maven-resolved dependency JARs, and the JDK, the index stores its fully-qualified name (FQN), simple name, package, accessible constructors, accessible methods, fields, and declared imports when source code is available. Thus, for each simple type name that may appear in a generated test, the ClassIndex maps the name to candidate FQNs drawn from three sources: (i)~project classes, including \texttt{src/main/java} and \texttt{src/test/java}, (ii)~dependency classes discovered from Maven-resolved JARs obtained with \texttt{mvn dependency:build-classpath}, and (iii)~JDK classes from a static JDK table. The index filters candidates by classpath visibility and Java accessibility constraints, then ranks the remaining candidates by package proximity to the CUT, explicit imports in the CUT, and source priority, preferring project-local symbols over external-library symbols. During repair, the \textsc{Fixer} queries the ClassIndex to identify unresolved types, invalid constructor invocations, nonexistent method calls, and inconsistent imports, then proposes replacements from the ranked candidate set.

\subsubsection{Typestate Model} For stateful classes whose public methods must be invoked in a particular order to avoid exceptions, we construct a Markov chain over method transitions. States represent the last method invoked on an object (i.e., the API's current usage context), with a distinguished $\texttt{\_\_INIT\_\_}$ state representing the initial state before any method call. A directed edge $m \rightarrow m'$ means that method $m'$ is a valid successor of method $m$, i.e., $m'$ can be invoked immediately after $m$ without violating the API usage constraints. For a current state $m$, the model assigns transition probabilities over all candidate successor methods $m''$ whose transitions from $m$ have not been blocked:
\begin{equation}
    P(m \!\to\! m') \;=\; \frac{\mathds{1}[(m,m') \notin \mathcal{B}]}{\lvert\{m'' : (m,m'') \notin \mathcal{B}\}\rvert},
    \label{eq:typestate}
\end{equation}
where $\mathcal{B}$ is the set of \textit{blocked} transitions accumulated at runtime, and $m''$ ranges over candidate successor methods of $m$. A blocked transition is a method pair $(m,m')$ observed to violate the object's usage protocol, for example, because invoking $m'$ immediately after $m$ causes an \texttt{IllegalStateException} or another state-related failure. The model is initialized \textit{statically} from source-code evidence before test generation: we parse the CUT and its reachable dependency usages with tree-sitter, group method calls by receiver object, and add an edge between two methods when they appear consecutively on the same receiver in an observed usage sequence. For example, a sequence such as \texttt{writer.setNextName(q); writer.writeStartObject();} creates the transition \texttt{setNextName} $\rightarrow$ \texttt{writeStartObject}. We also inspect field-guarded preconditions (e.g.\ ``\texttt{if \_nextName==null throw IllegalStateException}'') to infer required predecessor methods and to mark transitions that would violate the guard. At initialization, $\mathcal{B}$ is empty, and all inferred transitions receive equal probability mass, normalized by out-degree. The chain is then updated \textit{dynamically} from test outcomes (Section~\ref{sec:c-fixer}): every passing test reinforces the observed edges, and every state-related failure contributes a newly blocked edge to~$\mathcal{B}$, after which probabilities are renormalized over the remaining valid successors per Equation~(\ref{eq:typestate}).

\begin{figure}[hbt!]
    \centering
    \includegraphics[width=0.5\linewidth]{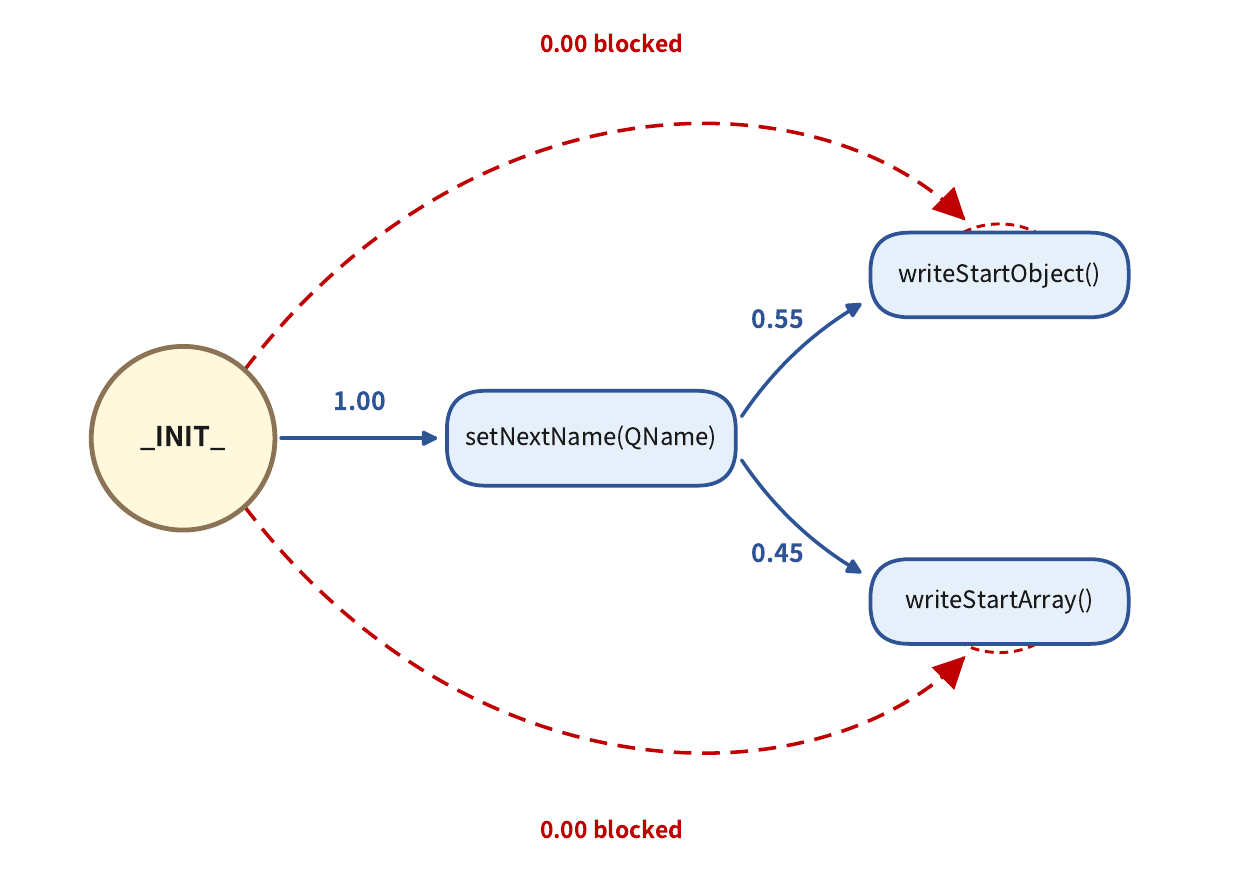}
    \caption{Example of a typestate model for JacksonXML.}
    \label{fig:typestate}
\end{figure}

Figure~\ref{fig:typestate} illustrates a typestate model derived from the motivating example in Section~\ref{sec:example}. The solid edges from \texttt{setNextName(QName)} to the two \texttt{writeStartX()} methods represent valid method transitions, with probabilities assigned according to the out-degree normalization in Equation~(\ref{eq:typestate}). In contrast, the dashed red edges from \texttt{\_\_INIT\_\_} directly to the \texttt{writeStartX()} methods represent blocked transitions in $\mathcal{B}$. These transitions are extracted statically from the precondition \texttt{if (\_nextName == null) throw IllegalStateException}, which indicates that a name must be set before any start-write operation is invoked. Therefore, any candidate test that calls a \texttt{writeStartX()} method before \texttt{setNextName(QName)} is flagged by the \textsc{Fixer} as a protocol-level violation, as described in Section~\ref{sec:c-fixer}.

The typestate model is used to guide generation rather than as an authoritative specification of program correctness. Because it is derived from potentially incomplete call sequences, guarded preconditions, and execution outcomes, it may exclude valid sequences or admit invalid ones. In particular, a failure caused by an implementation defect could be mistaken for evidence of an invalid transition, potentially steering subsequent generation away from a bug-revealing sequence. To mitigate this risk, we deliberately treat the model as a soft filter: a flagged transition triggers re-generation of the offending segment under protocol-level guidance rather than outright test rejection, and the model is refined dynamically from execution outcomes as described above. Empirically, the protocol-level ablation (Section~\ref{sec:rq3_results}) shows that removing this component degrades line coverage by 11.05 percentage points on \dfj and 14.69 on \dfpj, indicating that its benefits outweigh the risk of over-blocking in practice.

\subsection{Stage II: Iterative Plan--Generate--Validate--Fix Loop}
\label{sec:loop}

As shown in Figure~\ref{fig:overview}, this stage iteratively generates and repairs tests through a multi-agent workflow, progressively producing, validating, and refining test cases until it meets the termination criteria. We describe the agents in the order they execute within each loop iteration. Throughout this paper, we use the term \textit{agent} to denote a structured role with a dedicated prompt template and responsibility within the pipeline; each agent is implemented as a single LLM invocation with no tool use, persistent memory, or autonomous planning beyond its defined prompt. This design choice keeps the pipeline lightweight and reproducible while still benefiting from task decomposition.

\subsubsection{\textsc{Initializer} (Optional)}
\label{sec:c-init}
The \textsc{Initializer} runs once before the loop starts: it parses the CUT source with tree-sitter and emits a minimal test file containing the correct package declaration, the imports for the CUT and for JUnit, and an empty \texttt{@Test} placeholder per public CUT method. This ``test skeleton'' avoids wasting the first generation round on boilerplate and guarantees that every subsequent iteration edits a syntactically valid compilation unit. The step is optional. If an existing test file already exists (e.g.\ when \method is bootstrapped from a partial suite), \method appends new test cases instead of creating new test files.

\subsubsection{\textsc{Planner}}
\label{sec:c-planner}

The \textsc{Planner} turns the current coverage gap into a concrete \textit{test plan}. It builds a control-flow graph for every method of the CUT using COMEX~\cite{das2023comex}, a tree-sitter-based code-view generation tool that extracts structural program representations such as control-flow and data-flow graphs, and then enumerates the resulting paths. To identify the most-uncovered methods, we compute an \emph{uncoveredness score} for each method $m$ as $U(m) = 1 - \frac{|C(m)|}{|B(m)|}$, where $B(m)$ is the set of branches in $m$'s control-flow graph and $C(m) \subseteq B(m)$ is the subset covered by the current test suite, as reported by the latest JaCoCo execution. Methods are then ranked by $U(m)$ (ties broken by line coverage). Following the practice in PANTA~\cite{panta} and to prevent overly long prompts, we restrict each iteration to a small path budget: we pick the top two paths from each of the two highest-ranked uncovered methods (exploitation) and two paths from two randomly sampled methods (exploration), yielding at most $K{=}4$ target paths per iteration. Isomorphic paths (identical node sequences up to branch-merge reordering) are deduplicated before scoring, so that duplicated paths do not consume the budget.

Figure~\ref{fig:planner} shows the prompt template used by the \textsc{Planner}. The prompt provides the selected uncovered paths, the line-numbered class under test, and the current test file. Based on this information, the \textsc{Planner} generates 2--6 test plans, each targeting uncovered behavior in the selected paths. Each plan specifies the target method, the intended testing strategy, the required setup steps, and a short explanation of how the plan is expected to improve coverage.

\subsubsection{\textsc{Generator}}
\label{sec:c-generator}
\begin{figure}[t]
    \centering
    \subfloat[Prompt template for \textsc{Planner}.\label{fig:planner}]{%
        \includegraphics[width=0.48\linewidth]{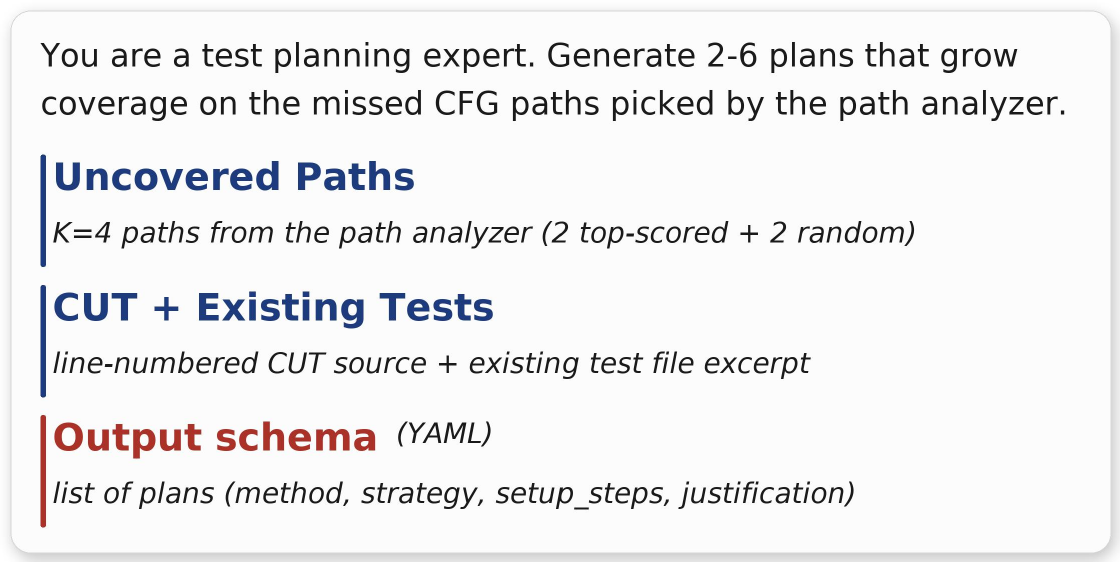}%
    }\hfill
    \subfloat[Prompt template for \textsc{Generator}.\label{fig:generator}]{%
        \includegraphics[width=0.48\linewidth]{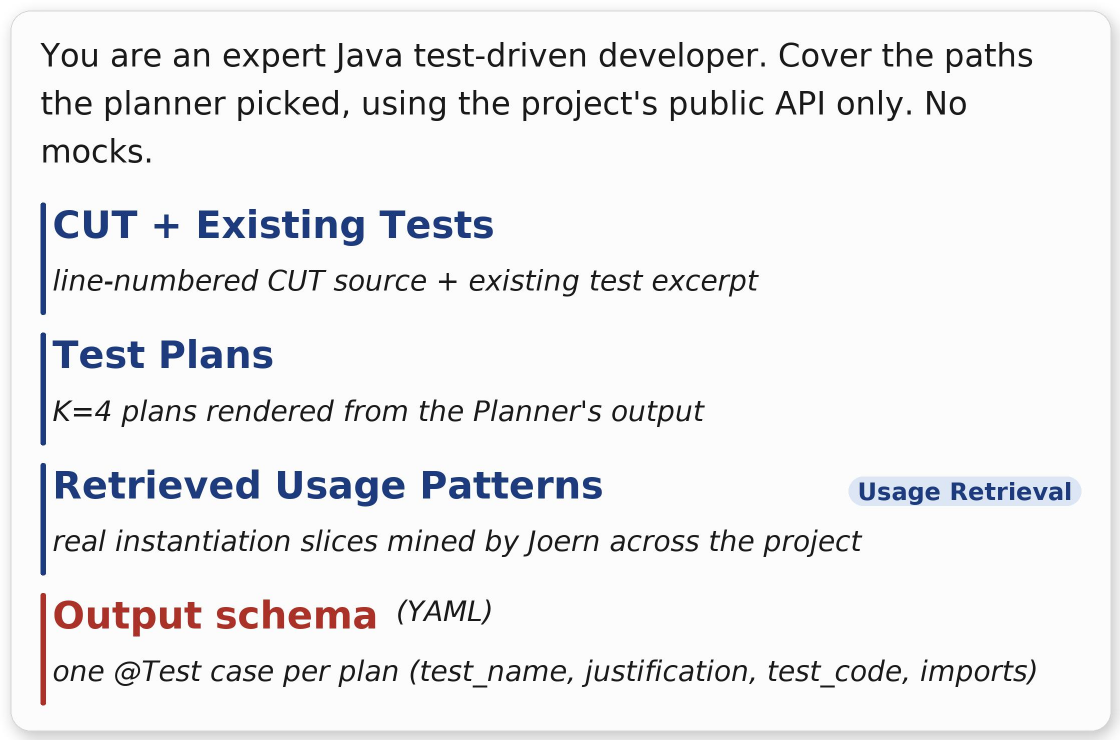}%
    }
    \caption{Prompt templates used by the planning and generation agents.}
    \label{fig:prompt-templates}
\end{figure}
The \textsc{Generator} produces a candidate test. In a naive baseline, it only receives the CUT source code and the test plan, which often leads to hallucinated dependencies and incorrect instantiation patterns (the \emph{``not knowing''} hallucination). \method addresses this by augmenting the Generator with a program slicer built on Joern, which enriches the prompt with real project usage patterns of both the CUT and its dependencies.

Given a dependency class $D$ appearing in the CUT’s signatures (constructor parameters, public-method parameters, fields, or return types), the slicer issues Joern queries to locate all call sites involving $D$ across both the project’s production and test code. For each call site, we perform backward slicing on the CPG to recover the \emph{minimal instantiation chain}: the shortest sequence of statements (measured by number of statements) that starts from variables available at the call site and ends with a fully constructed, type-correct object of $D$, including local variable definitions, factory invocations, and required imports. Minimality is defined over statement count; when multiple chains of equal length exist, the one whose first statement is lexically closest to the call site is preferred. The resulting slices are deduplicated using structural hashing. They are then ranked by lexicographic composition of three signals: (i) occurrence in passing tests, (ii) occurrence in production code, and (iii) simplicity of the instantiation chain (favoring shorter sequences), with signal (i) as the primary key, (ii) as the secondary key, and (iii) as the tiebreaker. The top-$k$ slices are then rendered as executable Java code snippets and injected into the Generator prompt to guide correct dependency instantiation.

Figure~\ref{fig:generator} shows the prompt used by the \textsc{Generator}. In addition to the class under test and the current test file, the prompt includes the test plans produced by the \textsc{Planner} and the retrieved usage patterns mined by the Joern-based slicer. These usage patterns provide concrete examples of how project classes are instantiated and used in existing code. The \textsc{Generator} uses them to produce one \texttt{@Test} method for each plan, together with the required imports and a brief explanation. By supplying real dependency usage patterns, this prompt helps reduce errors caused by missing project-specific context (``not knowing''), such as hallucinated constructors, factory methods, or invalid object setup.

\subsubsection{\textsc{Validator}}
\label{sec:c-validator}
The \textsc{Validator} takes each candidate test produced by the Generator and determines whether it passes, fails, or should be forwarded to the Fixer. Concretely, it (i) compiles the test together with the existing test suite against the project’s Maven-resolved classpath, (ii) executes it under JUnit with a per-test timeout, and (iii) analyzes the resulting compile-time and runtime diagnostics to produce a structured error report. This report is then forwarded to the Fixer. Tests that compile, execute, and pass are added to the accumulated test suite and used to update the typestate model via Equation~(\ref{eq:typestate}).

\subsubsection{\textsc{Fixer}}
\label{sec:c-fixer}
The \textsc{Fixer} addresses the second class of limitations (\emph{``not following''}), which occur when the model fails to adhere to required constraints even when they are explicitly provided. Instead of relying on a single repair step, the Fixer performs a two-stage correction process for each failing test. The process starts with the failing test and its compiled error signals from execution and compilation. \method first generates an initial repair based on this feedback (\textit{Fixer I}). The resulting test is then checked against three constraint levels: (i) symbol-level constraints from the ClassIndex, which ground the repair in project-visible classes, constructors, methods, and imports before any higher-level reasoning is applied;  (ii) protocol-level constraints from the Markov typestate model that enforce valid API call ordering; and (iii) iteration-level guidance from experience memory, which contains gold tests, fix recipes, and anti-patterns. Gold tests are previously generated tests that compile and pass, fix recipes summarize successful edits that repaired earlier failures, and anti-patterns record unfixable tests or repair attempts that should not be repeated. If any violation is detected, a second repair is triggered, conditioned on the identified constraint violations (\textsc{Fixer II}). In this second repair, the model must also produce a structured justification explaining why the revised test satisfies all constraints and resolves the original failure. This is a key design choice in our approach: by forcing the model to explicitly explain its own correction, we make constraint satisfaction part of the generation process rather than an implicit expectation. The justification requires the model to reflect on symbol-, protocol-, and iteration-level constraints, strengthening adherence during repair.

\noindent\textbf{Symbol-level constraints.} Symbol-level failures, including fabricated classes, invalid method calls, incorrect constructor signatures, and illegal abstract instantiations, occur when the model produces identifiers that do not exist in the project. To eliminate such errors, the Fixer performs deterministic symbol verification using the ClassIndex.

For each failing test, all referenced symbols are resolved against the ClassIndex, and invalid references are replaced with valid candidates. Method calls are checked against the index entries, and mismatches are replaced with the closest valid alternative based on string similarity or removed if no safe replacement exists. Constructor invocations are similarly validated against the recorded constructor signatures, ensuring arity and type consistency.

Abstract or interface instantiations are replaced with concrete implementations retrieved from the ClassIndex based on type compatibility and package proximity. Missing or ambiguous imports are resolved using the same index, with project-local definitions prioritized over external dependencies.

\noindent\textbf{Protocol-level constraints.} Protocol-level failures arise when the generated test violates the required API usage order, such as invoking methods before required initialization steps. We address this by validating the test against the Markov typestate model.

Each method call is mapped to a state transition. Invalid transitions, defined as transitions with zero probability or as transitions along previously blocked edges, identify the first point of violation in the execution sequence. The \textsc{Fixer} then reconstructs a valid call ordering by replacing the offending segment with a feasible sequence that satisfies the typestate constraints. For example, consider a generated test that invokes \texttt{writeStartObject()} before calling \texttt{setNextName(QName)} on a streaming writer. According to the typestate model, the transition from the initial state directly to \texttt{writeStartObject()} is invalid because it omits the required intermediate state induced by \texttt{setNextName}. The Fixer identifies this violation and reconstructs the sequence by inserting the required initialization call, resulting in a valid ordering: \texttt{setNextName(QName)} followed by \texttt{writeStartObject()}.

\noindent\textbf{Iteration-level constraints.} The ClassIndex and typestate model define \emph{what is valid}, but do not capture \emph{which repair strategies are effective in practice}. Experience memory addresses this gap by retaining a structured record of past repair outcomes.

\begin{compactitem}
    \item \textit{Successful repairs (gold memory):} Each successful fix is stored as a structured triple consisting of the error signature, the applied correction, and a before/after code diff. When a similar error recurs, the system retrieves the most similar prior case and injects it into the Fixer prompt as a concrete repair example.

    \item \textit{Failed attempts:} Persistently failing patterns are stored as anti-patterns and fed back to the Generator as negative guidance, reducing repetition of known-invalid structures.

    \item \textit{Unfixable cases:} Errors that exhaust $N_{\mathrm{fix}}$ repair attempts are recorded as unfixable patterns and optionally used to update the typestate constraints when applicable, preventing repeated exploration of infeasible behaviors.
\end{compactitem}

\begin{figure}[t]
    \centering
    \subfloat[Prompt template for \textsc{Fixer I}.\label{fig:fixer1}]{%
        \includegraphics[width=0.48\linewidth]{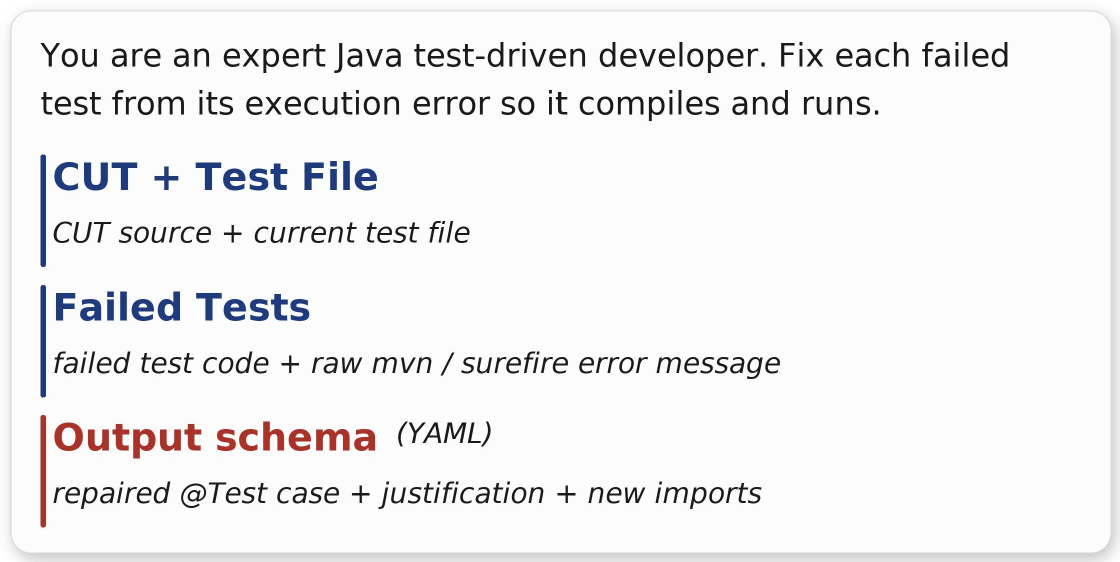}%
    }\hfill
    \subfloat[Prompt template for \textsc{Fixer II}.\label{fig:fixer2}]{%
        \includegraphics[width=0.48\linewidth]{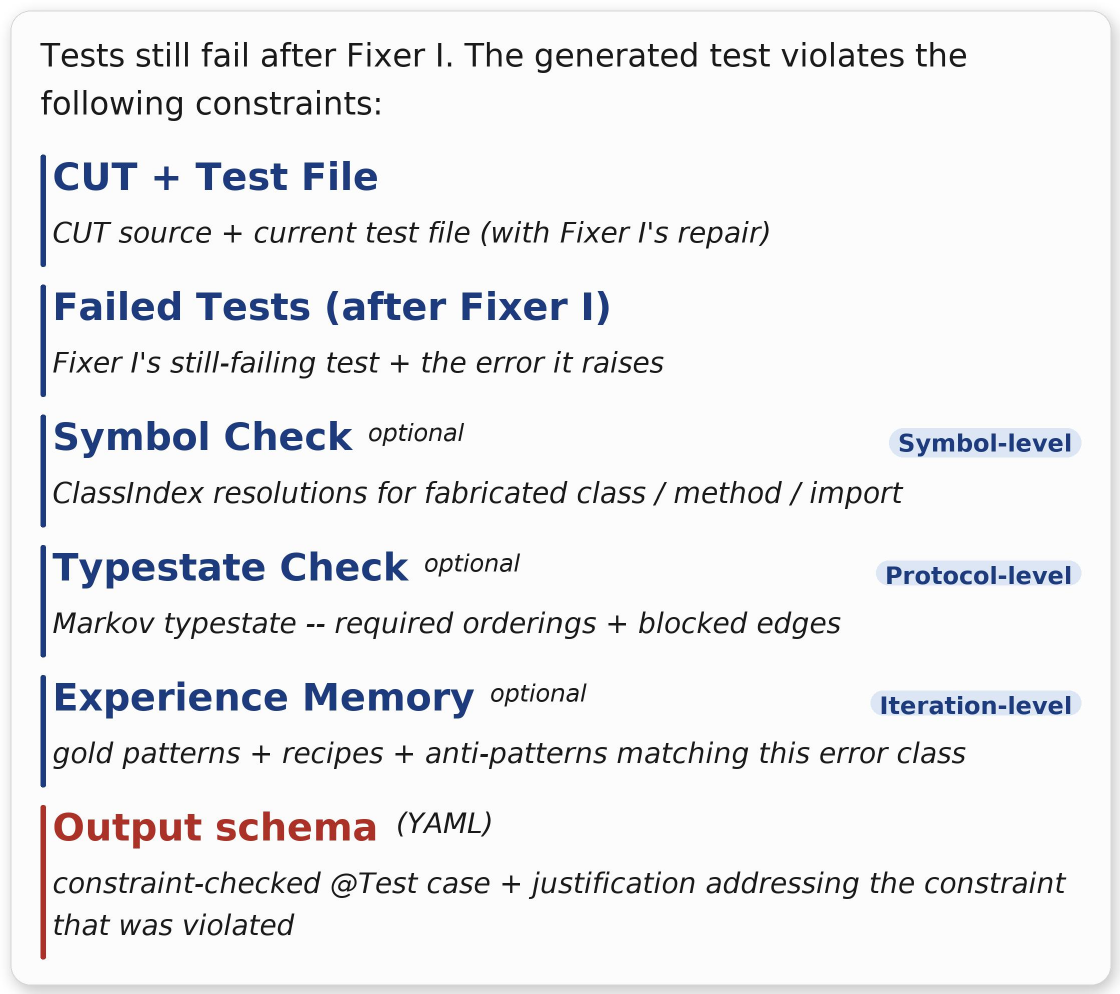}%
    }
    \caption{Prompt templates used by the two fixing stages.}
    \label{fig:fixer-prompts}
\end{figure}

Overall, the three components operate at complementary levels: the ClassIndex enforces correctness at the symbol level, the typestate model enforces correctness at the protocol level, and experience memory improves repair efficiency across iterations. Together, they shift constraint satisfaction from a purely language-driven process to a machine-enforced correction pipeline.
Figure~\ref{fig:fixer1} shows the prompt template for \textsc{Fixer I}. This prompt provides the class under test, the current test file, and the diagnostics for failing tests, including compilation errors or Surefire execution failures. The goal of this stage is to repair simple failures directly from the observed error messages. The model returns a revised \texttt{@Test} method, any required imports, and a short explanation of the change. This first repair stage does not include symbol, typestate, or memory constraints; it defers these checks to the second stage so trivial errors can be fixed with a lightweight prompt.
 
Figure~\ref{fig:fixer2} shows the second repair prompt, which is used when the first repair still violates project-specific constraints. In addition to the failed test and its diagnostics, this prompt may include three types of constraint information. The symbol check reports invalid or ambiguous classes, methods, constructors, or imports identified using the ClassIndex. The typestate check reports invalid call orders, required method sequences, and blocked transitions from the Markov typestate model. The experience memory provides relevant successful repairs and known anti-patterns from previous iterations. Using these inputs, the model generates a constraint-checked test and explains how the revised test addresses the reported violations. This makes the repair process explicitly guided by project-specific constraints rather than relying only on the model's interpretation of error messages.

\section{Experimental Design}
\label{sec:design}
\subsection{Research Questions}
\label{sec:rqs} 
We structure our empirical evaluation around three research questions.

\begin{compactitem}
    \item \textbf{RQ1 (Effectiveness):} How effective is \method in generating low-level integration tests for Java repositories compared with the LLM-based and search-based baselines?  

    \item \textbf{RQ2 (Efficiency):} What efficiency cost does \method incur, in terms of wall-clock time and token consumption, relative to the baselines? 

    \item \textbf{RQ3 (Ablation):} How much do the major strategies in \method, namely \textit{usage retrieval}, \textit{iteration-level memory}, \textit{protocol constraints}, and \textit{symbol constraints}, each contribute to its effectiveness?  
\end{compactitem}

For RQ1, we compare \method against two baselines: PANTA, the SOTA LLM-based approach, and EvoSuite, the SOTA search-based tool, to assess whether LLM-based integration test generation can improve test effectiveness compared to unit-level generation. This comparison examines both the quality of valid tests produced and the extent to which the generated tests exercise production code, including the CUT and code reached through real dependency interactions. Section~\ref{sec:metrics} introduces the detailed effectiveness metrics.

For RQ2, we analyze the practical cost of applying \method. Since integration test generation requires LLMs to reason about real dependencies, construct valid object states, and iteratively repair generated tests, it may introduce additional overhead compared to unit-level generation approaches. We therefore evaluate the computational cost of test generation in time and token consumption, with detailed efficiency metrics in Section~\ref{sec:metrics}.

For RQ3, we conduct an ablation study to quantify how the major components of \method support its two core strategies: context-enriched generation and constraint-enforced fixing. We compare the full approach with four variants. The first variant, \textit{No\_UR}, removes usage retrieval and therefore directly ablates the context-enriched generation strategy by providing no mined dependency-usage examples to the generator. The other three variants ablate the constraint-enforced fixing strategy at different levels: \textit{No\_Sym} removes symbol-level constraints, so the repair stage is no longer explicitly restricted to valid project-visible classes, methods, and fields; \textit{No\_Prot} removes protocol-level constraints, allowing the model to violate valid API call sequences and object-state transitions; and \textit{No\_Iter} disables iteration-level constraints, i.e., experience memory, so information from gold tests, successful fix recipes, and unfixable anti-patterns is no longer carried into later repair iterations. These variants are chosen because they isolate the retrieval component used to address \emph{not knowing} and the three complementary constraint levels used to address \emph{not following}.
\subsection{Datasets} 
\label{sec:datasets}
We evaluate \method on two complementary benchmarks: \dfj, a widely used Java test-generation benchmark that enables direct comparison with prior work, and \dfpj, a new, more challenging benchmark curated from post-cutoff open-source projects. While \dfj provides a standard basis for evaluation, it may suffer from potential data leakage and relies on older Java and JUnit versions. To address these limitations, we construct \dfpj to (i) reduce the risk of data leakage, (ii) evaluate whether \method generalizes across diverse projects and dependency structures, and (iii) assess its effectiveness under more modern Java environments and more complex multi-module repository scenarios. Table~\ref{tab:bench-compare} summarizes the main characteristics of the two benchmarks, including the number of projects (``\#Projects''), selected classes (``\#Classes'') and methods (``\#MUTs''), average cyclomatic complexity (``CC''), Java and JUnit versions, and the earliest project creation time (``Created After'').

\begin{table}[hbt!]
\centering
\caption{Characteristics of \dfj and \dfpj.}
\label{tab:bench-compare}
\small
\setlength{\tabcolsep}{5pt}
\resizebox{0.8\columnwidth}{!}{
\begin{tabular}{cccccccc}
\toprule
Benchmark & \#Projects & \#Classes & \#MUTs & CC & Java & JUnit & Created After\\
\midrule
\dfj   & 14 & 130 & 2971 & 16.5 & 11 & 4 & 2008 \\
\dfpj  &  5 &  30 & 540 & 38.4 & 17 & 5 & 2025 \\
\bottomrule
\end{tabular}
}
\end{table}

\noindent\textbf{\dfj.} Following recent research~\cite{panta,Hossain2024TOGLL:LLMs}, we use the \dfj~v2.0.0 bug-free revisions of 14 Java projects. We focus on the 130 non-trivial classes identified by Gu et al.~\cite{panta} as having cyclomatic complexity (CC) $\geq 11$ (i.e., we exclude simple classes that admit trivial single-path tests). These projects span Apache Commons (\texttt{Lang}, \texttt{Math}, \texttt{Csv}, \texttt{Cli}, \texttt{Codec}, \texttt{Collections}, \texttt{Compress}), Jackson (\texttt{JacksonCore}, \texttt{JacksonDatabind}, \texttt{JacksonXml}), \texttt{Jsoup}, \texttt{JxPath}, \texttt{Gson}, and \texttt{JodaTime}. All projects target Java~11 and  JUnit~4. Their public commit dates range from 2008 to 2018, preceding the release date of any LLM considered in this study. We therefore cannot rule out potential data leakage in this dataset.  

\begin{table}[hbt!]
\centering
\caption{Statistics of the \dfpj dataset.}
\label{tab:deps4j-projects}
\footnotesize
\setlength{\tabcolsep}{4pt}
\resizebox{0.8\columnwidth}{!}{
\begin{tabular}{ccccc}
\toprule
Project  & Domain & \#CUTs & \#MUTs & CC \\
\midrule
adk-java \cite{adk-java}             & Agent Development Kit (Google)   & 6 & 118 & 38.7 \\
fit-framework \cite{fit-framework}   & Model-engineering utils (Huawei) & 6 &  89 & 17.8 \\
agentscope-java \cite{agentscope}    & Multi-agent runtime              & 6 & 174 & 87.0 \\
a2a-java \cite{a2a-java}             & Agent-to-Agent SDK (Google)      & 8 & 107 & 29.0 \\
jquick-curl \cite{jquick-curl}       & ANTLR-generated curl parser      & 4 &  52 & 14.8 \\
\midrule
\textbf{Total}                       &                                  & \textbf{30} & \textbf{540} & \textbf{38.4} \\
\bottomrule
\end{tabular}
}
\end{table}

\noindent\textbf{\dfpj.} To mitigate the data leakage threat inherent in \dfj, and to evaluate \method on software that reflects modern Java practice, we curate \dfpj: a new benchmark of 30 classes drawn from 5 high-profile open-source projects whose first commit dates and complete development histories post-date the training-data cutoff of the LLM used in our experiments (Qwen3-Coder, cutoff 2025-03). The selected projects (see Table~\ref{tab:deps4j-projects}) include \texttt{adk-java} (Google Agent Development Kit), \texttt{fit-framework} (Huawei Model Engine), \texttt{agentscope-java}, \texttt{a2a-java} (Google Agent-to-Agent SDK), and \texttt{jquick-curl}. We selected these projects because they (i)~have been public after the year 2025, (ii)~target Java~17 with JUnit~5, exercising new language features that were unavailable in the \dfj projects, and (iii)~are actively starred and forked on GitHub, indicating engineering quality comparable to \dfj subjects. Among Java 17 projects created after the cutoff, these five were chosen to maximize diversity in dependency structures: they span agent runtimes (adk-java, a2a-java, agentscope-java), general-purpose utilities (fit-framework), and generated parsers (jquick-curl), with cyclomatic complexity ranging from 14.8 to 87.0 (Table~\ref{tab:deps4j-projects}) and with different Maven module layouts, ensuring that the benchmark is not biased toward a single project archetype.

\subsection{Baselines}
\label{subsec:baseline}
We compare \method against two baselines representing the two main families of automated test generation: \textbf{PANTA}~\cite{panta}, the state of the art among LLM-based approaches for Java, and \textbf{EvoSuite}~\cite{evosuite2014}, the state of the art among traditional search-based approaches~\cite{Fraser2011EvoSuite:Software,evosuite2014}.

\noindent\textit{PANTA} is the strongest available LLM-based baseline for our setting. It is the most recent LLM-based approach that handles classes with external dependencies and establishes SOTA results on the widely adopted \dfj benchmark. \noindent\textit{EvoSuite} is the most widely used and strongest search-based test generation tool for Java and implements the DynaMOSA~\cite{dynamosa} algorithm by default. We choose EvoSuite over the alternative random-based tool Randoop~\cite{Pacheco2007Randoop:Java} because comparative studies consistently show EvoSuite achieving higher coverage and fault detection~\cite{evosuite2014}. 

\noindent\textbf{Baseline Selection Criteria.} We therefore focus on PANTA and EvoSuite, which jointly cover the two dominant paradigms (LLM-based and search-based) and are both directly applicable to our benchmarks with publicly available implementations. Other approaches like CANDOR~\cite{candor} target method-level programs without external dependencies (e.g., HumanEvalJava) and cannot generate tests for the dependency-rich classes in our benchmarks. Aster~\cite{pan2025aster} and CITYWALK~\cite{zhang2025citywalk} target Python and C++, respectively, which are not directly applicable to Java projects. System-level tools such as EvoMaster~\cite{evomaster} require REST API specifications, while our benchmarks consist of library and framework code without such exposed endpoints. 
\subsection{Evaluation Metrics \& Statistical Testing}
\label{sec:metrics}
We evaluate generated test suites from two perspectives: \textit{effectiveness} and \textit{efficiency}. Effectiveness metrics measure how well the generated tests exercise the production code and expose behavioral differences. Efficiency metrics quantify the computational cost of producing the tests, including time, token consumption, and effort for iterative refinement. Finally, we use statistical testing to account for randomness in LLM-based generation and assess whether observed differences between approaches are statistically significant.

\subsubsection{Effectiveness metrics}
\label{sec:metrics-quality}
We adopt the following metrics to evaluate the effectiveness:

\noindent\textit{Number of Tests.} This metric reports the number of valid test cases in the final generated test file. We include this metric to quantify how many executable tests each approach produces. In unit test generation, generated tests may be discarded because they fail to compile, fail during execution, violate framework constraints, or rely on invalid dependency interactions.  Thus, more valid tests indicate that the approach can produce more runnable tests during generation and repair.

\noindent\textit{Line and Branch Coverage.} We record standard JaCoCo line coverage and branch coverage on the CUT. These two metrics quantify how thoroughly the suite exercises the CUT in isolation and are directly comparable to the numbers reported by prior LLM-based generators~\cite{panta,candor}.

\noindent\textit{Mutation Score.} Line and branch coverage measure \emph{reach} but not \emph{sensitivity}: a test that invokes a method without asserting anything meaningful will still increase coverage. We therefore also report the mutation score computed with PIT~\cite{pit} using the default operator set. For a CUT with $N$ generated mutants, the score is the fraction that are killed by at least one test in the generated suite: $\mathit{MS}=\lvert\text{killed}\rvert/N$.

\noindent\textit{Dependency Line Coverage (DepLC).}
Line and branch coverage are computed \emph{only on the CUT}. An integration test suite, however, also exercises the CUT's real dependencies. The central motivation is that these dependency interactions create a coverage gap left by unit-level testing approaches (Section~\ref{sec:intro}). To quantify this effect, we extend JaCoCo's class-level reporting to the entire Maven module containing the CUT, and define three derived metrics over the resulting coverage vector:
\begin{align}
\mathit{DLC}    &= \sum_{\ell \in \mathit{CUT}} \mathds{1} [\ell\text{ covered}], \label{eq:dlc} \\
\mathit{TLC}    &= \sum_{\ell \in \mathit{Module}} \mathds{1}[\ell\text{ covered}], \label{eq:tlc} \\
\mathit{DepLC}  &= \mathit{TLC} - \mathit{DLC}, \label{eq:deplc} 
\end{align}
where $\mathit{DLC}$ (Direct Line Coverage) counts the covered lines in the CUT itself, $\mathit{TLC}$ (Transitive Line Coverage) counts the covered lines across all production classes in the CUT's Maven module, and $\mathit{DepLC}$ isolates the additional covered lines outside the CUT. Thus, $\mathit{DepLC}$ captures the extent to which generated tests execute real project code beyond the target class. This complements the CUT-level coverage and mutation score: while those metrics evaluate how thoroughly the target class is tested in isolation, $\mathit{DepLC}$ measures the additional dependency code exercised through integration-level execution.

\subsubsection{Efficiency metrics}
\label{sec:metrics-efficiency}
We adopt the following metrics to evaluate the efficiency:

\noindent\textit{Number of Iterations.} This metric reports the number of \texttt{plan-generate-validate-fix} cycles executed for a CUT before the tool stops, either by reaching its iteration budget, satisfying the stopping criterion, or failing to improve further. We include this metric to measure each approach's convergence efficiency.
 
\noindent\textit{Tokens Per Method.} This metric reports the average token consumption per method in the CUT. We include this metric to measure each approach's cost efficiency at the method level. Token consumption directly reflects the computational and monetary cost of producing tests. Normalizing token usage by the number of methods in the CUT makes the comparison less sensitive to differences in class or project size, and provides a clearer view of how much LLM budget is required to test each unit of code. 

\noindent\textit{Tokens Per Iteration.}  This metric is calculated as the total token consumption divided by the number of iterations. We include this metric to measure each approach's cost efficiency per iteration.  In iterative test generation, higher coverage often requires additional iterations, increasing total token consumption. Therefore, a larger total token count may reflect more extensive refinement rather than lower efficiency. By normalizing token consumption per iteration, this metric captures the average token cost per generation cycle and provides a fairer comparison of per-iteration efficiency across approaches.

\noindent\textit{Time Per Method.}
This metric is calculated as the total wall-clock time divided by the number of methods in the CUT, where the total time is the elapsed time from the start to the end of test generation for the CUT. We include this metric to measure the time required to test each unit of code. Since different CUTs may contain different numbers of methods, the total time alone can be strongly affected by target size and may not provide a fair comparison across CUTs or projects. By normalizing wall-clock time per method, this metric provides a clearer view of average generation cost per method and reflects an approach's practical scalability on larger code units.

\noindent\textit{Time Per Iteration.}
This metric is the total wall-clock time divided by the number of generate--validate--fix iterations, where the total time is the elapsed time from the start to the end of test generation for a CUT. We include this metric because total time alone does not distinguish between approaches that perform different numbers of cycles. In iterative test generation, additional iterations may be necessary to improve the test suite and reach higher coverage, which naturally increases total runtime. Therefore, a larger total time may reflect more extensive refinement rather than lower efficiency. By normalizing time to the number of iterations, this metric captures the average time cost per cycle and enables a fairer comparison of per-iteration efficiency across approaches.

\subsubsection{Statistical Testing}
\label{sec:statistical-testing}

LLMs are inherently stochastic, as different runs may produce different tests even under identical experimental settings. To mitigate randomness, we report average results on the \dfj (130 CUTs) and \dfpj (30 CUTs) datasets. We also conduct pairwise statistical testing and calculate effect sizes for each metric. 

Following the guidance for statistical testing~\cite{Arcuri2011AEngineering}, we use the two-sided paired Wilcoxon signed-rank test to compare \method against each baseline (PANTA and EvoSuite) on the per-CUT results, and report the Vargha--Delaney $\hat{A}_{12}$ effect size~\cite{a12}, which is widely used for randomized software-engineering algorithms~\cite{Arcuri2011AEngineering}. In our context, $\hat{A}_{12}$ estimates the probability that \method outperforms the baseline on a randomly selected paired observation, with ties counted as half. A value of $0.5$ indicates no preference between the two approaches; values above $0.5$ indicate that \method tends to achieve higher values, while values below $0.5$ indicate that the baseline tends to achieve higher values. Following Vargha and Delaney~\cite{a12}, we interpret the effect as \emph{negligible} when $|\hat{A}_{12}-0.5| < 0.06$, \emph{small} when $0.06 \leq |\hat{A}_{12}-0.5| < 0.14$, \emph{medium} when $0.14 \leq |\hat{A}_{12}-0.5| < 0.21$, and \emph{large} when $|\hat{A}_{12}-0.5| \geq 0.21$. For the \method--EvoSuite mutation comparison, the test is computed on the 119 CUTs for which EvoSuite produced measurable mutation scores. We count unmeasurable CUTs (instrumentation failures and timeouts) as zero in the reported averages but exclude them from the paired test to avoid biasing the comparison.

\subsection{Implementation Details}
\label{sec:impl}

All experiments run on a Precision~7960 Tower workstation with an Intel Xeon w9-3495X processor and dual NVIDIA RTX~6000 Ada GPUs. The implementation is written in Python and uses LangChain~\cite{2025LangchainWebsite} for LLM integration. We use Qwen3-Coder-30B-A3B-Instruct-FP8~\cite{qwen3coder} as the backbone model and serve it locally with vLLM~\cite{vllm}. This local deployment avoids API costs and privacy concerns associated with closed-source LLM services, making the experimental setting better suited to real projects.  

To ensure a fair comparison, we configure both \method and PANTA under identical settings. Specifically, both tools use a context window of \num{16384}  tokens, a sampling temperature of 0.2, and a maximum output length of \num{4096}  tokens, as in PANTA~\cite{panta}. We set the maximum number of generation iterations $N_{iter}$ to 30, the maximum number of repair attempts $ N _ {fix} $ to 5, and the plateau-based early-stopping patience to 4. We selected these values based on PANTA experimental results, where test generation for most CUTs converged under this configuration. Using the same generous budget lets us fully exercise both PANTA and \method and ensures the comparison is not biased by premature stopping. For both tools, we set the generation target to 100\% line coverage. Test generation stops once this target is reached or when the iteration budget or plateau criterion is triggered. For the program slicer used in \method, we adopt Joern, a widely used code analysis framework that supports code property graph-based analysis. For mutation testing, we use the default set of mutation operators, following the practice of MUTGEN~\cite{mutgen} and PANTA~\cite{panta}.

For EvoSuite, we use version 1.2.0 with its default DynaMOSA search. Each CUT is given a 300-second search budget, but CUTs that do not converge within this budget are re-run with an increased budget of up to one hour. All CUTs are thus complete by convergence (median 55 seconds), and EvoSuite's results are not limited by an insufficient budget. EvoSuite's mutation score is measured with PIT using the same operator set and flags as for \method and PANTA; CUTs where EvoSuite produced no usable tests or where PIT timed out are counted as zero (details in Section~\ref{sec:threats}).  On Deps4J (Java 17), EvoSuite 1.2.0 cannot process Java 17 bytecode and completes for only 4 of 30 CUTs (all in jquick-curl). We therefore report EvoSuite results on \dfj in full and only its jquick-curl results on \dfpj, which we discuss separately in Section~\ref{sec:rq1_results}.
\section{Experimental Results}
\label{sec:results}
\subsection{RQ1 Results (Effectiveness)}
\label{sec:rq1_results}
\begin{table*}[t]
    \centering
    \scriptsize
    \caption{Comparison of IntTestGen ($M$), Panta ($P$), and EvoSuite ($E$) on Defects4J and Deps4J. \textit{Number of Tests}: total tests per project (EvoSuite: sum over CUTs of the mean tests per successful seed). \textit{Line/Branch Coverage}: per-CUT averages. \textit{\#Mutants}: total mutants per project, identical across systems since PIT generates the same mutant set. \textit{Mutation Score}: percentage of killed mutants. \textit{Project DepLC}: covered dependency lines, project-level union. The \textit{Stat.~test} rows report the paired Wilcoxon $p$-value and $\hat{A}_{12}$ effect size of IntTestGen against each baseline: values under $P$ are M vs.\ P; values under $E$ are M vs.\ E (per-CUT tests for line/branch/mutation; per-project tests for test count and DepLC). ``--'': not available or not comparable.}
    \label{tab:rq1}
    \resizebox{\linewidth}{!}{%
    \begin{tabular}{ll ccc ccc ccc c ccc ccc}
    \toprule
    \multirow{2}{*}{\textbf{Dataset}} & \multirow{2}{*}{\textbf{Project}} & \multicolumn{3}{c}{\textbf{Number of Tests}} & \multicolumn{3}{c}{\textbf{Line Coverage}} & \multicolumn{3}{c}{\textbf{Branch Coverage}} & \multirow{2}{*}{\textbf{\#Mutants}} & \multicolumn{3}{c}{\textbf{Mutation Score}} & \multicolumn{3}{c}{\textbf{Project DepLC}} \\
    \cmidrule(lr){3-5}\cmidrule(lr){6-8}\cmidrule(lr){9-11}\cmidrule(lr){13-15}\cmidrule(lr){16-18}
    & & $M$ & $P$ & $E$ & $M$ & $P$ & $E$ & $M$ & $P$ & $E$ & & $M$ & $P$ & $E$ & $M$ & $P$ & $E$ \\
    \midrule
    \multirow{15}{*}{Defects4J} & Cli & 65 & 75 & 40.8 & 96.18 & 78.13 & \textbf{96.2} & 89.88 & 57.44 & \textbf{95.8} & 173 & 57.23 & 43.93 & \textbf{57.9} & 138 & 131 & \textbf{211} \\
     & Codec & 196 & 284 & 200.8 & 90.95 & 83.92 & \textbf{97.4} & 86.57 & 74.29 & \textbf{96.4} & 1207 & 37.37 & 40.51 & \textbf{66.9} & 415 & 420 & \textbf{429}  \\
     & Collections & 387 & 351 & 351.3 & \textbf{93.41} & 70.79 & 80.4 & \textbf{90.60} & 67.75 & 81.1 & 711 & \textbf{72.86} & 57.81 & 58.1 & 717 & 781 & \textbf{1189}  \\
     & Compress & 197 & 524 & 125.3 & \textbf{85.89} & 55.77 & 48.2 & \textbf{78.77} & 49.43 & 47.7 & 1570 & \textbf{40.76} & 24.01 & 25.9 & \textbf{2368} & 17 & 691  \\
     & Csv & 142 & 254 & 135.8 & \textbf{96.52} & 69.40 & 89.4 & 87.90 & 55.25 & \textbf{89.1} & 419 & 29.36 & 21.72 & \textbf{64.4} & \textbf{84} & 69 & 76  \\
     & Gson & 162 & 619 & 185.4 & \textbf{86.86} & 74.98 & 80.2 & \textbf{81.96} & 60.81 & 77.5 & 771 & \textbf{56.55} & 53.96 & 48.2 & 587 & 776 & \textbf{1045}  \\
     & JCore & 394 & 486 & 606.6 & 80.49 & 51.02 & \textbf{87.5} & 74.69 & 44.79 & \textbf{86.3} & 3364 & 34.42 & 14.57 & \textbf{50.8} & 1311 & 620 & \textbf{1396}  \\
     & JDatabind & 417 & 770 & 41.7 & \textbf{81.64} & 55.29 & 9.4 & \textbf{76.10} & 46.74 & 8.9 & 1443 & \textbf{60.22} & 41.72 & 4.8 & \textbf{3133} & 1833 & 0  \\
     & JXml & 219 & 97 & 0.0 & \textbf{80.23} & 64.83 & 0.0 & \textbf{80.12} & 54.14 & 0.0 & 652 & \textbf{47.09} & 23.47 & 0.0 & \textbf{221} & 21 & 0  \\
     & Jsoup & 254 & 593 & 297.0 & 89.76 & 78.30 & \textbf{91.9} & 80.42 & 63.35 & \textbf{89.9} & 979 & \textbf{58.84} & 37.18 & 35.7 & 1256 & 1150 & \textbf{1448}  \\
     & JxPath & 426 & 584 & 127.5 & \textbf{84.45} & 52.29 & 34.5 & \textbf{80.27} & 45.33 & 33.9 & 1351 & \textbf{58.48} & 50.85 & 21.0 & \textbf{589} & 349 & 77  \\
     & Lang & 1153 & 2281 & 1620.4 & 86.39 & 71.12 & \textbf{88.9} & 83.00 & 64.22 & \textbf{85.7} & 4209 & 49.06 & 27.89 & \textbf{56.0} & 352 & 352 & \textbf{492}  \\
     & Math & 1075 & 2244 & 993.7 & \textbf{92.72} & 74.09 & 68.9 & \textbf{89.10} & 65.50 & 67.9 & 8155 & \textbf{48.95} & 35.27 & 35.8 & \textbf{3410} & 2902 & 2726  \\
     & Time & 552 & 632 & 40.6 & \textbf{97.94} & 83.63 & 9.0 & \textbf{92.94} & 74.67 & 9.0 & 1348 & \textbf{76.85} & 63.72 & 6.3 & \textbf{2172} & 2047 & 0  \\
    \cmidrule(lr){2-18}
     & \textbf{Average} & 402.79 & 699.57 & 340.5 & \textbf{88.82} & 68.83 & 61.52 & \textbf{83.74} & 58.84 & 60.42 & 1882.3 & \textbf{52.00} & 38.33 & 35.59 & \textbf{1197} & 819 & 699 \\
    \midrule
    \multirow{6}{*}{Deps4J} & A2A & 92 & 225 & -- & \textbf{87.04} & 69.31 & -- & \textbf{59.35} & 55.39 & -- & 214 & 43.46 & \textbf{59.35} & -- & \textbf{58} & 1 & -- \\
     & ADK & 123 & 257 & -- & 76.25 & \textbf{78.25} & -- & \textbf{61.31} & 60.96 & -- & 298 & 48.32 & \textbf{55.03} & -- & 101 & \textbf{145} & -- \\
     & AgentScope & 173 & 226 & -- & \textbf{76.52} & 44.76 & -- & \textbf{69.02} & 33.83 & -- & 551 & \textbf{54.45} & 43.19 & -- & \textbf{717} & 534 & -- \\
     & FIT & 112 & 237 & -- & \textbf{79.76} & 66.06 & -- & \textbf{70.62} & 57.90 & -- & 298 & \textbf{79.19} & 74.83 & -- & \textbf{490} & 411 & -- \\
     & JQuick & 72 & 9 & 208.3 & 60.34 & 8.07 & \textbf{69.1} & 30.29 & 3.62 & \textbf{58.7} & 547 & \textbf{10.05} & 2.19 & -- & 27 & 27 & \textbf{76} \\
    \cmidrule(lr){2-18}
     & \textbf{Average} & 114.4 & 190.8 & 41.7 & \textbf{75.98} & 53.29 & 9.2 & \textbf{58.12} & 42.34 & 7.8 & 381.6 & \textbf{47.09} & 46.92 & -- & \textbf{279} & 224 & 15.2 \\
    \midrule
    \multirow{2}{*}{\textbf{Stat.\ test}} & p-value & - & $<0.001$ & $0.187$ & - & $<0.001$ & $<0.001$ & - & $<0.001$ & $<0.001$ & -- & - & $0.001$ & $<0.001$ & - & $0.010$ & $0.315$ \\
     & $\hat{A}_{12}$ & - & 0.158 & 0.679 & - & 0.947 & 0.65 & - & 1.000 & 0.59 & -- & - & 0.842 & 0.70 & - & 0.737 & 0.658 \\
    \bottomrule
    \end{tabular}%
    }
\end{table*}

Table~\ref{tab:rq1} reports the three-way comparison of \method, PANTA, and EvoSuite on \dfj and \dfpj. On the dataset level, \method achieves the highest line coverage (88.82\%), followed by PANTA (68.83\%) and EvoSuite (61.52\%); the same ranking holds for mutation score (52.00\% vs.\ 38.33\% vs.\ 35.59\%). For branch coverage, \method (83.74\%) leads both PANTA (58.84\%) and EvoSuite (60.42\%). The Wilcoxon signed-rank tests indicate that all pooled improvements of \method over PANTA are statistically significant ($p<0.05$), with large effect sizes ($|\hat{A}_{12}-0.5|\geq 0.21$). For \method over EvoSuite, paired Wilcoxon tests on all 130 CUTs are also significant for line coverage ($p{=}4.0{\times}10^{-8}$, $\hat{A}_{12}{=}0.65$), branch coverage ($p{=}6.6{\times}10^{-6}$, $\hat{A}_{12}{=}0.59$), and mutation score over the 119 CUTs with measurable EvoSuite scores ($p{=}1.5{\times}10^{-8}$, $\hat{A}_{12}{=}0.70$).  

\noindent \textbf{PANTA Comparison. }On \dfj, \method generates 297 fewer tests per project on average than PANTA (402.79 vs.\ 699.57), while achieving higher line and branch coverage on every project. In particular, \method improves average line coverage by \SI{19.99}{\pp} and average branch coverage by \SI{24.90}{\pp}. Even its lowest project-level coverage remains high, with \SI{80.23}{\percent} line coverage on JXml and \SI{74.69}{\percent} branch coverage on JCore, suggesting that \method is practically effective for real-world Java integration testing. \method also achieves higher mutation scores on 13 of the 14 \dfj{} projects, with Codec as the only exception ($\Delta=-\SI{3.14}{\pp}$). On average, it improves mutation score by \SI{13.67}{\pp}, indicating that the generated tests are not only broad in coverage but also strong at fault detection. Finally,
\method increases project DepLC by 378 lines on average (1197 vs.\ 819), consistent with its design to exercise real dependencies: by executing dependency code rather than mocked surrogates, it exercises substantially more non-CUT code.

We observe similar patterns on \dfpj. \method again generates fewer tests per project on average than PANTA (114.4 vs.\ 190.8), while improving average line coverage by \SI{22.69}{\pp} and average branch coverage by \SI{15.78}{\pp}. The overall mutation scores are nearly identical (\SI{47.09}{\percent} for \method vs.\ \SI{46.92}{\percent} for PANTA). At the project level, \method performs better on AgentScope, FIT, and JQuick, while PANTA performs better on A2A and ADK. One reason is that PANTA generates substantially more tests on \dfpj, giving it more opportunities to introduce assertions and kill mutants. This gives PANTA an inherent advantage in mutation score, which depends not only on exercised code but also on the number and strength of test oracles. In addition, \dfpj contains more complex projects, many of which are multi-module and involve richer dependency interactions. While exercising real dependencies helps \method execute more non-CUT code, producing precise assertions for such complex behaviors remains challenging. Thus, achieving a comparable mutation score with fewer tests suggests that \method provides competitive fault-detection capability while substantially improving line, branch, and dependency coverage. Finally, \method improves project DepLC by an average of 55 lines (279 vs.\ 224). The consistent improvements on both \dfj and \dfpj suggest that \method is not overfit to a specific benchmark, but generalizes across Java repositories from different domains. Moreover, because the \dfpj projects were created after Qwen3-Coder's training cutoff date, the improvements are unlikely to be due to data leakage; rather, they are attributable to \method's design.

Particularly, both \method and PANTA achieve their worst overall performance on JQuick. For \method, line coverage and branch coverage on JQuick are only \SI{60.34}{\percent} and \SI{30.29}{\percent}, respectively, substantially lower than its \dfpj averages of \SI{75.98}{\percent} and \SI{58.12}{\percent}. PANTA shows an even sharper drop, with line and branch coverage decreasing to \SI{8.07}{\percent} and \SI{3.62}{\percent}, respectively. The mutation score and DepLC also reach their lowest values among all 19 projects. This result is surprising because JQuick has the lowest cyclomatic complexity among the five \dfpj projects, as shown in Table~\ref{tab:deps4j-projects}. JQuick is a Java HTTP client framework that translates cURL commands into executable HTTP requests, allowing developers to issue requests without manually constructing low-level HTTP client code.  Effective testing of such parser-centric code requires inputs that satisfy the underlying cURL grammar. However, JQuick's source code and its dependencies do not include this grammar. As a result, both PANTA and \method often generate malformed or uninformative cURL strings. Moreover, parser feedback is usually coarse-grained, such as generic parse failures, and offers little actionable guidance on which grammar rule was violated or how to revise the input. Consequently, the \textit{plan--generate--validate--fix} loop becomes much less effective than it is for other projects, where failures often expose clear exceptions, return values, or state changes. This explains why both tools perform poorly on JQuick despite its relatively low cyclomatic complexity.

This finding suggests an important direction for future work: combining \method with grammar-aware or fuzzing-assisted input generation. For parser-centric components, fuzzers can systematically explore valid and near-valid inputs, while \method can use the resulting executions, failures, and seed inputs to construct valid test cases and assertions. Such a hybrid design could make the feedback loop more informative for grammar-driven code.

\noindent\textbf{EvoSuite comparison.} EvoSuite's results on \dfj are highly uneven across projects: it matches or exceeds \method on utility-heavy projects (Codec: 97.4\% vs.\ 90.95\%; JCore: 87.5\% vs.\ 80.49\%; Lang: 88.9\% vs.\ 86.39\%), but collapses on protocol-heavy and dependency-dense projects (JDatabind: 9.4\%; Time: 9.0\%; JXml: 0.0\% ). This pattern is expected, since EvoSuite's search optimizes local coverage objectives and cannot infer the project-specific API protocols that \method mines through program slicing and enforces through typestate constraints. On the classes where EvoSuite succeeds, its mutation score is competitive (e.g., Cli 57.9\%, Codec 66.9\%). The overall gap stems from dependency-rich classes that require integration-level reasoning. In terms of DepLC, \method covers 1{,}197 dependency lines per project on average versus 819 for PANTA and 699 for EvoSuite (Table~\ref{tab:rq1}).  Without using any mocks, \method covers more dependency lines and achieves a higher mutation score (52.00\% vs.\ 38.33\% and 35.59\%) than both baselines, exercising real dependency code rather than simulated surrogates and showing that real-dependency execution delivers both broader reach and stronger fault detection. On Deps4J, EvoSuite completes only the four jquick-curl CUTs, and the reason is instructive: EvoSuite is a fixed tool whose bytecode instrumentation cannot handle Java 17 class files, so it fails on the 26 CUTs from projects that use modern language features (e.g., records and sealed classes). LLM-based approaches do not have this limitation: they generate source-level JUnit code and can therefore target any Java version, which is why IntTestGen and PANTA run on all 30 Deps4J CUTs. This flexibility matters in practice, as low-level integration testing of modern codebases increasingly requires covering code that uses new language features.

\noindent\textbf{Detailed Analysis on DepLC.}
DepLC is one of the central claims of this work: an integration test suite should exercise substantially more real dependency code than a unit-level suite.  To verify that DepLC results reflect genuine integration-level execution rather than a measurement artifact, we further examine where dependency coverage comes from in each system: how much the baselines replace real dependencies with mocks, and how dependency coverage behaves on the exact CUTs where mocking is used.

\noindent\textit{Prevalence of mocking in the baselines.}
We scan the generated test files of both baselines for Mockito API usage (\texttt{mock()}, \texttt{when()}, \texttt{verify()}, and related annotations). On \dfj, 7 of 130 PANTA CUTs (5.4\%) contain Mockito-based tests, with approximately 555 \texttt{mock()} calls across the benchmark. EvoSuite relies on mocking substantially more: 36 of the 101 CUTs  (35.3\%) contain Mockito-based tests, with 1,177 \texttt{mock()} calls, and a further 16.9\% of its successful runs use EvoSuite's runtime \emph{environment} mocks (which replace JDK classes such as files and system calls). On \dfpj, PANTA uses Mockito in 6 of 25 comparable CUTs (24\%), reflecting the more complex dependency structures of multi-module projects. These are concentrated in a2a-java (3 CUTs), where agent-to-agent SDK classes require constructing deeply nested dependency graphs.

\noindent\textit{Mock intensity within affected CUTs.}
Within PANTA's mock-using CUTs, we measure the proportion of \texttt{@Test} methods that invoke at least one Mockito API. On \dfj, these CUTs average 68\% of test methods containing mocking calls; on \dfpj, 73\%. For example, on \texttt{ToXmlGenerator}, 7 of 9 PANTA-generated tests use mocks, while \method generates tests exercising the real Jackson XML writer. When PANTA encounters complex dependencies, mocking is its dominant strategy, systematically trading dependency coverage for testability.

\noindent\textit{DepLC gaps on mock-heavy CUTs.}
For the 7 mock-heavy, comparable CUTs on \dfj (excluding JxPath due to hard-linked test paths), \method covers 749 more dependency lines than PANTA. The largest gaps occur on protocol-heavy data-format classes: \texttt{ToXmlGenerator} (+129), \texttt{FromXmlParser} (+156), and \texttt{TokenFilterContext} (+404). These are precisely the classes where correct dependency construction and API call ordering are critical---the challenges that mocking sidesteps but \method's constraint-enforced fixing is designed to address. On a2a-java's three mock-heavy CUTs, the gap is smaller (+49), not because mocking is harmless, but because these classes involve external service interactions that are inherently difficult to exercise even with real dependencies.

Taken together, these findings verify the central claim behind DepLC: the gap is not a measurement artifact but reflects genuine integration-level execution. \method exercises real dependencies, while the baselines' reliance on mocking leaves substantial dependency code unexecuted. This finding aligns with our core motivation: unit-level approaches that default to mocking create a measurable blind spot that low-level integration testing can systematically address.

\begin{center}
\fbox{\parbox{0.97\linewidth}{
\textbf{Answer to RQ1:} \method is effective in generating low-level integration tests for Java repositories, outperforming both the LLM-based baseline PANTA and the search-based baseline EvoSuite in line coverage, branch coverage, and mutation score on \dfj, with statistically significant differences (Wilcoxon $p<0.05$ for both comparisons). On \dfpj, \method also outperforms PANTA, while EvoSuite cannot process most Java 17 projects.
}}
\end{center}

\subsection{RQ2 Results (Efficiency)}
\label{sec:rq2_results}
\begin{table*}[t]
    \centering
    \tiny
    \caption{Comparison of MocklessTester and Panta on cost / efficiency metrics. Each metric spans three columns: MocklessTester ($M$), Panta ($P$), and $\Delta = M - P$. Token-based columns are reported in thousands. Tokens-per-method and time-per-method normalise cost by output volume; the per-iteration columns normalise by the number of generations. The bottom \textit{Stat.~test} row reports the two-sided paired Wilcoxon signed-rank \textit{p-value} comparing $M$ against $P$ across all projects and the Vargha--Delaney $\hat{A}_{12}$ effect size: $\hat{A}_{12}>0.5$ indicates $M$ uses more, $<0.5$ indicates $P$ uses more. ``--'' denotes not applicable.}
    \label{tab:rq2}
    \resizebox{\linewidth}{!}{%
    \begin{tabular}{ll*{5}{rr>{\columncolor{gray!12}}r}}
    \toprule
    \multirow{2}{*}{\textbf{Dataset}} & \multirow{2}{*}{\textbf{Project}} & \multicolumn{3}{c}{\textbf{Mean Iterations}} & \multicolumn{3}{c}{\textbf{Tokens / Method (k)}} & \multicolumn{3}{c}{\textbf{Tokens / Iteration (k)}} & \multicolumn{3}{c}{\textbf{Time / Method (s)}} & \multicolumn{3}{c}{\textbf{Time / Iteration (s)}} \\
    \cmidrule(lr){3-5}\cmidrule(lr){6-8}\cmidrule(lr){9-11}\cmidrule(lr){12-14}\cmidrule(lr){15-17}
    & & $M$ & $P$ & $\Delta$ & $M$ & $P$ & $\Delta$ & $M$ & $P$ & $\Delta$ & $M$ & $P$ & $\Delta$ & $M$ & $P$ & $\Delta$ \\
    \midrule
    \multirow{15}{*}{\dfj} & Cli & 15 & 2.5 & +12.50 & 27.90 & 2.81 & +25.10 & 60.45 & 42.09 & +18.37 & 109.25 & 192.02 & -82.77 & 236.71 & 2880.33 & -2643.62 \\
     & Codec & 14 & 9.86 & +4.14 & 19.10 & 10.42 & +8.68 & 38.20 & 42.88 & -4.68 & 95.31 & 55.21 & +40.10 & 190.61 & 227.25 & -36.64 \\
     & Collections & 19.2 & 4 & +15.20 & 10.81 & 2.10 & +8.71 & 43.57 & 36.90 & +6.67 & 69.85 & 35.32 & +34.53 & 281.58 & 619.78 & -338.20 \\
     & Compress & 15.67 & 9 & +6.67 & 51.46 & 11.74 & +39.72 & 71.90 & 75.95 & -4.06 & 221.63 & 45.12 & +176.51 & 309.65 & 291.90 & +17.75 \\
     & Csv & 20 & 8 & +12.00 & 30.95 & 4.84 & +26.11 & 73.25 & 51.24 & +22.01 & 116.59 & 15.52 & +101.07 & 275.93 & 164.25 & +111.68 \\
     & Gson & 15.75 & 10.5 & +5.25 & 28.56 & 5.10 & +23.46 & 73.44 & 75.13 & -1.69 & 108.39 & 12.43 & +95.96 & 278.71 & 183.21 & +95.50 \\
     & JCore & 18.67 & 6.33 & +12.34 & 26.75 & 5.78 & +20.96 & 62.72 & 49.31 & +13.41 & 101.32 & 51.54 & +49.78 & 237.63 & 439.41 & -201.78 \\
     & JDatabind & 19.33 & 7.56 & +11.77 & 30.72 & 5.39 & +25.34 & 73.63 & 61.01 & +12.62 & 91.58 & 32.01 & +59.57 & 219.48 & 362.44 & -142.96 \\
     & JXml & 22.75 & 3.75 & +19.00 & 31.70 & 5.32 & +26.38 & 76.28 & 34.37 & +41.91 & 101.50 & 209.17 & -107.67 & 244.27 & 1352.65 & -1108.38 \\
     & Jsoup & 15.5 & 4.62 & +10.88 & 26.36 & 2.38 & +23.98 & 54.00 & 38.14 & +15.85 & 124.18 & 48.41 & +75.77 & 254.37 & 775.94 & -521.57 \\
     & JxPath & 17.08 & 9.92 & +7.16 & 34.85 & 11.59 & +23.26 & 72.43 & 56.88 & +15.55 & 123.38 & 57.11 & +66.27 & 256.39 & 280.28 & -23.89 \\
     & Lang & 21.29 & 8.94 & +12.35 & 15.92 & 3.83 & +12.09 & 50.71 & 57.50 & -6.79 & 68.80 & 12.77 & +56.03 & 219.12 & 191.65 & +27.47 \\
     & Math & 13.7 & 6.4 & +7.30 & 20.67 & 3.16 & +17.50 & 54.05 & 36.97 & +17.08 & 120.74 & 26.73 & +94.01 & 315.81 & 312.42 & +3.39 \\
     & Time & 15 & 7.64 & +7.36 & 16.50 & 5.85 & +10.65 & 55.20 & 44.03 & +11.16 & 72.99 & 19.34 & +53.65 & 244.18 & 145.54 & +98.64 \\
    \cmidrule(lr){2-17}
     & \textbf{Average} & 17.35 & 7.07 & +10.28 & 26.59 & 5.74 & +20.85 & 61.42 & 50.17 & +11.25 & 108.97 & 58.05 & +50.92 & 254.60 & 587.65 & -333.05 \\
    \midrule
    \multirow{6}{*}{\dfpj} & A2A & 6.75 & 6.25 & +0.50 & 28.34 & 12.86 & +15.48 & 48.29 & 57.89 & -9.61 & 81.06 & 67.16 & +13.90 & 138.10 & 302.22 & -164.12 \\
     & ADK & 9.83 & 9.67 & +0.16 & 31.48 & 17.36 & +14.12 & 65.62 & 76.91 & -11.29 & 86.37 & 67.41 & +18.96 & 180.06 & 298.71 & -118.65 \\
     & AgentScope & 15.33 & 6.33 & +9.00 & 38.00 & 9.70 & +28.30 & 71.46 & 57.71 & +13.75 & 90.90 & 38.33 & +52.57 & 170.92 & 227.95 & -57.03 \\
     & FIT & 5.83 & 6.4 & -0.57 & 17.04 & 6.49 & +10.55 & 54.53 & 48.08 & +6.45 & 50.34 & 28.17 & +22.17 & 161.10 & 208.60 & -47.50 \\
     & JQuick & 8.5 & 2 & +6.50 & 12.45 & 2.25 & +10.19 & 26.36 & 10.13 & +16.23 & 40.58 & 16.76 & +23.82 & 85.93 & 75.44 & +10.49 \\
    \cmidrule(lr){2-17}
     & \textbf{Average} & 9.25 & 6.13 & +3.12 & 25.46 & 9.73 & +15.73 & 53.25 & 50.14 & +3.11 & 69.85 & 43.57 & +26.28 & 147.22 & 222.58 & -75.36 \\
    \midrule
    \multirow{2}{*}{\textbf{Stat. test}} & \textit{p-value} & -- & -- & $<$.001 & -- & -- & $<$.001 & -- & -- & 0.005 & -- & -- & 0.009 & -- & -- & 0.029 \\
     & $\hat{A}_{12}$ & -- & -- & 0.947 & -- & -- & 1.000 & -- & -- & 0.684 & -- & -- & 0.895 & -- & -- & 0.368 \\
    \bottomrule
    \end{tabular}%
    }
\end{table*}

Table~\ref{tab:rq2} compares the efficiency of \method and PANTA. Overall, \method consumes more total LLM tokens and wall-clock time than PANTA. This overhead is expected because \method continues generating for substantially more iterations than PANTA, which tends to stop earlier. The Wilcoxon signed-rank tests show that the differences between \method and PANTA are statistically significant for all efficiency metrics (\textit{p-value}$\leq0.05$), with large effect sizes ($|\hat{A}_{12}-0.5|>0.21$), except for \textit{Tokens/Iteration}, where the effect size is small ($\hat{A}_{12}=0.684$).


On \dfj, \method consumes 26.59k tokens per method on average and takes 108.97 seconds per method. In comparison, PANTA consumes 5.74k tokens per method and takes 58.05 seconds. Although both tools use the same iteration budget (30 iterations with plateau-based early-stopping patience of 4), PANTA stops much earlier on average. This suggests that \ method's higher total cost mainly comes from its longer generation process, which lets it keep improving coverage after PANTA has plateaued. When normalizing by the number of iterations, the token gap becomes much smaller: \method consumes only 11.25k more tokens per iteration than PANTA. More importantly, \method is substantially faster per iteration, requiring 254.60 seconds per iteration compared with 587.65 seconds for PANTA, a reduction of approximately \SI{58.7}{\percent}. This suggests that although \method performs more iterations, each iteration is more time-efficient.

The results on \dfpj show a similar trend. \method consumes more tokens and time per method than PANTA, with 25.46k vs.\ 9.73k tokens and 69.85 vs.\ 43.57 seconds, respectively. However, after accounting for the number of iterations, the token gap becomes much smaller again: \method requires only 3.11k more tokens per iteration than PANTA. However, at the iteration level, \method requires approximately \SI{33.9}{\percent} less time per iteration than PANTA.

These results show that \method incurs additional total cost mainly because it performs more iterations to pursue higher coverage, rather than because each iteration is inefficient. In absolute terms, the cost remains practical: \method requires less than two minutes per method on average and stays well below 60k tokens per method. Based on a conservative estimate using public API prices from providers such as OpenRouter and Alibaba, generating tests for the entire \dfj benchmark across all 14 projects would cost only about \$12--\$45. Such overhead is acceptable for integration test generation, especially considering the substantial effectiveness gains reported in RQ1. Moreover, the wall-clock time can be further reduced by using more powerful GPUs or commercial API-based LLM services.

\noindent\textbf{EvoSuite efficiency.} EvoSuite is not included in Table~\ref{tab:rq2} because its cost is fundamentally different: as a search-based tool, it consumes CPU time but no LLM tokens, so token-based metrics are inapplicable, and its wall-clock time is not directly comparable to LLM-based generation (which includes inference on a GPU workstation). We therefore report its efficiency here for completeness. We gave EvoSuite a 300-second search budget per CUT, increasing it to up to one hour for CUTs that did not converge within that time. All reported runs terminate by convergence, and budget constraints do not limit their results.  In short, EvoSuite is the cheapest tool in generation time but the least effective in coverage (RQ1). \method trades additional inference cost for substantially higher effectiveness, and this trade-off remains practical at less than two minutes per method.

\begin{center}
\fbox{\parbox{0.97\linewidth}{\textbf{Answer to RQ2:}
\method incurs a higher total token and time cost than PANTA because it performs more iterations to achieve higher coverage. However, its per-iteration token cost is competitive, and its per-iteration runtime is substantially lower than PANTA's. In absolute terms, \method remains practical, requiring less than two minutes and fewer than 60k tokens per method on average.}}
\end{center}
\subsection{RQ3 Results (Ablation Study)}
\label{sec:rq3_results}
\begin{table*}[t]
    \centering
    \scriptsize
    \caption{Ablation study: full \method (\textit{M}) vs four ablations -- \textit{-UR} removes usage pattern retrieval, \textit{-Sym} removes symbol-level contraints, \textit{-Prot} removes protocol-level contraints, \textit{-Iter} removes iteration-level contraints. Each metric spans five columns; bold marks the value that beats every other variant within that project. Project DepLC is the count of distinct lines covered in classes outside the CUT. The bottom \textit{Stat.~test} row reports the two-sided paired Wilcoxon signed-rank \textit{p-value} comparing \textit{M} against each variant across all projects and the Vargha--Delaney $\hat{A}_{12}$ effect size. ``--'' denotes not applicable.}
    \label{tab:rq3}
    \resizebox{\linewidth}{!}{%
    \begin{tabular}{ll *{5}{c} *{5}{c} *{5}{c} *{5}{c}}
    \toprule
    \multirow{2}{*}{\textbf{Dataset}} & \multirow{2}{*}{\textbf{Project}}
    & \multicolumn{5}{c}{\textbf{Line \%}}
    & \multicolumn{5}{c}{\textbf{Branch \%}}
    & \multicolumn{5}{c}{\textbf{Mutation Score\%}}
    & \multicolumn{5}{c}{\textbf{Project DepLC}} \\
    \cmidrule(lr){3-7}\cmidrule(lr){8-12}\cmidrule(lr){13-17}\cmidrule(lr){18-22}
    & & M & -UR & -Sym & -Prot & -Iter
    & M & -UR & -Sym & -Prot & -Iter
    & M & -UR & -Sym & -Prot & -Iter
    & M & -UR & -Sym & -Prot & -Iter \\
    \midrule
    \multirow{15}{*}{\dfj} & Cli & 96.18 & 89.75 & \textbf{99.29} & 97.12 & 97.36 & 89.88 & 77.16 & 91.93 & \textbf{94.80} & 93.97 & 57.23 & 34.10 & \textbf{65.32} & 55.49 & 61.27 & \textbf{138} & 89 & 84 & 107 & 84 \\
     & Codec & 90.95 & 93.82 & 94.22 & \textbf{94.92} & 94.15 & 86.57 & 89.12 & 89.13 & 90.00 & \textbf{90.13} & 37.37 & \textbf{49.38} & 48.96 & 41.34 & 41.59 & \textbf{415} & 325 & 331 & 329 & 331 \\
     & Collections & \textbf{93.41} & 74.92 & 76.50 & 71.84 & 73.32 & \textbf{90.60} & 72.97 & 73.14 & 68.04 & 68.27 & \textbf{72.86} & 51.90 & 54.57 & 46.98 & 48.66 & \textbf{717} & 470 & 518 & 473 & 506 \\
     & Compress & \textbf{85.89} & 75.42 & 82.26 & 70.23 & 72.36 & \textbf{78.77} & 67.50 & 71.40 & 61.69 & 61.96 & \textbf{40.76} & 36.24 & 32.36 & 34.33 & 31.02 & \textbf{2368} & 1814 & 1707 & 1672 & 1559 \\
     & Csv & \textbf{96.52} & 94.21 & 93.42 & 90.95 & 86.87 & \textbf{87.90} & 86.57 & 84.38 & 82.42 & 77.46 & \textbf{29.36} & 27.45 & 24.58 & 26.49 & 25.06 & \textbf{84} & 70 & 71 & 67 & 65 \\
     & Gson & 86.86 & 85.35 & \textbf{89.35} & 75.60 & 87.52 & 81.96 & 83.94 & \textbf{87.59} & 72.15 & 79.44 & \textbf{56.55} & 53.31 & 55.25 & 31.13 & 52.92 & 587 & 542 & 569 & \textbf{679} & 646 \\
     & JCore & \textbf{80.49} & 70.16 & 58.10 & 62.55 & 70.05 & \textbf{74.69} & 61.49 & 51.34 & 54.88 & 61.69 & \textbf{34.42} & 24.64 & 15.07 & 21.52 & 21.70 & \textbf{1311} & 770 & 563 & 876 & 908 \\
     & JDatabind & \textbf{81.64} & 68.89 & 67.84 & 62.51 & 70.01 & \textbf{76.10} & 63.30 & 65.38 & 59.81 & 65.50 & \textbf{60.22} & 39.57 & 39.99 & 31.74 & 42.00 & \textbf{3133} & 2604 & 1905 & 1421 & 2735 \\
     & JXml & \textbf{80.23} & 55.70 & 55.09 & 49.28 & 61.48 & \textbf{80.12} & 44.81 & 50.86 & 43.82 & 54.97 & \textbf{47.09} & 22.55 & 23.77 & 21.78 & 22.55 & \textbf{221} & 158 & 181 & 161 & 170 \\
     & Jsoup & 89.76 & 90.59 & 93.47 & \textbf{93.48} & 92.10 & 80.42 & 77.56 & 82.08 & \textbf{84.80} & 81.48 & 58.84 & 54.65 & \textbf{59.96} & 57.20 & 58.43 & 1256 & 1331 & 1374 & \textbf{1437} & 1329 \\
     & JxPath & \textbf{84.45} & 67.79 & 78.66 & 78.50 & 80.12 & \textbf{80.27} & 62.99 & 73.62 & 73.65 & 75.82 & \textbf{58.48} & 51.30 & 52.85 & 55.14 & 55.00 & 589 & 176 & \textbf{1416} & 1378 & 1280 \\
     & Lang & \textbf{86.39} & 79.81 & 81.08 & 76.67 & 77.24 & \textbf{83.00} & 77.65 & 77.71 & 74.59 & 74.11 & \textbf{49.06} & 33.25 & 36.97 & 33.61 & 33.55 & 352 & 315 & \textbf{361} & 346 & 355 \\
     & Math & \textbf{92.72} & 78.37 & 75.96 & 76.55 & 79.18 & \textbf{89.10} & 71.94 & 70.13 & 70.57 & 73.39 & \textbf{48.95} & 32.88 & 33.66 & 31.97 & 30.03 & \textbf{3410} & 2261 & 2316 & 2141 & 2347 \\
     & Time & \textbf{97.94} & 81.36 & 84.47 & 88.52 & 89.60 & \textbf{92.94} & 74.54 & 79.58 & 84.34 & 84.95 & \textbf{76.85} & 54.90 & 60.22 & 65.95 & 64.17 & \textbf{2172} & 1714 & 1679 & 1761 & 1771 \\
    \cmidrule(lr){2-22}
     & \textbf{Average} & \textbf{88.82} & 79.01 & 80.69 & 77.77 & 80.81 & \textbf{83.74} & 72.25 & 74.88 & 72.54 & 74.51 & \textbf{52.00} & 40.44 & 43.11 & 39.62 & 42.00 & \textbf{1197} & 903 & 934 & 918 & 1006 \\
    \midrule
    \multirow{6}{*}{\dfpj} & A2A & \textbf{87.04} & 44.76 & 63.01 & 56.08 & 59.00 & \textbf{59.35} & 39.66 & 50.05 & 38.07 & 37.93 & 43.46 & 55.67 & \textbf{65.88} & 55.67 & 46.39 & 58 & 63 & 67 & 67 & 67 \\
     & ADK & 76.25 & 79.36 & \textbf{81.84} & 75.28 & 78.57 & 61.31 & 65.11 & \textbf{65.17} & 62.80 & 62.59 & 48.32 & 50.16 & \textbf{50.53} & 45.37 & 48.69 & 101 & 26 & \textbf{151} & 26 & 47 \\
     & AgentScope & \textbf{76.52} & 71.33 & 72.37 & 74.16 & 76.03 & \textbf{69.02} & 59.18 & 62.70 & 62.21 & 65.34 & \textbf{54.45} & 39.20 & 34.46 & 33.65 & 41.16 & \textbf{717} & 353 & 439 & 421 & 325 \\
     & FIT & 79.76 & 80.24 & \textbf{81.25} & 80.93 & 80.51 & 70.62 & 71.45 & 72.33 & \textbf{72.50} & 71.27 & \textbf{79.19} & 76.11 & 71.88 & 72.18 & 79.08 & \textbf{490} & 205 & 231 & 218 & 231 \\
     & JQuick & \textbf{60.34} & 13.60 & 25.39 & 20.00 & 35.61 & \textbf{30.29} & 7.99 & 18.86 & 14.89 & 29.47 & 10.05 & 0.00 & 0.00 & 0.00 & \textbf{14.54} & \textbf{27} & 0 & 0 & 0 & 13 \\
    \cmidrule(lr){2-22}
     & \textbf{Average} & \textbf{75.98} & 57.86 & 64.77 & 61.29 & 65.94 & \textbf{58.12} & 48.68 & 53.82 & 50.09 & 53.32 & \textbf{47.09} & 44.23 & 44.55 & 41.37 & 45.97 & \textbf{279} & 129 & 178 & 146 & 137 \\
    \midrule
    \multirow{2}{*}{\textbf{Stat. test}} & \textit{p-value} & -- & $<$.001 & 0.006 & $<$.001 & 0.005 & -- & $<$.001 & 0.003 & $<$.001 & 0.002 & -- & 0.002 & 0.023 & $<$.001 & 0.008 & -- & $<$.001 & 0.020 & 0.020 & 0.020 \\
     & $\hat{A}_{12}$ & -- & 0.789 & 0.684 & 0.789 & 0.684 & -- & 0.789 & 0.684 & 0.737 & 0.737 & -- & 0.842 & 0.737 & 0.895 & 0.737 & -- & 0.895 & 0.737 & 0.789 & 0.737 \\
    \bottomrule
    \end{tabular}%
    }
\end{table*}

Table~\ref{tab:rq3} reports the results of ablation studies, where we compare \method (denoted as M) with four ablated variants: \textit{-UR} (removing usage retrieval), \textit{-Sym} (removing symbol-level constraints), \textit{-Prot} (removing protocol-level constraints), and \textit{-Iter} (removing iteration-level constraints). By comparing \method across all four variants, we assess the individual contributions of usage retrieval, symbol-level constraints, protocol-level constraints, and iteration-level constraints to \method's effectiveness. Overall, we find that the full \method outperforms all variants in all metrics. The Wilcoxon signed-rank tests confirm the significance of all the gaps (\textit{p-value}$\leq0.05$) and the effect sizes are all large ($|\hat{A}_{12}-0.5|>0.21$), except for three medium effect sizes on line coverage of \textit{-Sym}, branch coverage of \textit{-Sym} and line coverage of \textit{-Sym}.  

\textit{Usage retrieval} provides usage patterns to help \method instantiate objects. When it is removed, average DepLC drops from 1197 to 903 on \dfj{} and from 279 to 129 on \dfpj{}, the largest reduction of any ablation on both datasets. It also causes broad declines in line, branch, and mutation score: on \dfj{}, average line coverage drops from 88.82 to 79.01, branch coverage from 83.74 to 72.25, and mutation score from 52.00 to 40.44; on \dfpj{}, the corresponding averages fall from 75.98 to 57.86, from 58.12 to 48.68, and from 47.09 to 44.23. This suggests that repository-specific usage examples do more than improve instantiation correctness: they also help the model generate tests that execute deeper behaviors once dependencies are reached.

\textit{Symbol-level constraints} have a more moderate but still important effect. At the dataset level, \textit{-Sym} lowers all four metrics on both benchmarks relative to the full method: on \dfj{}, line coverage falls from 88.82\% to 80.69\%, branch coverage from 83.74\% to 74.88\%, mutation score from 52.00\% to 43.11\%, and DepLC from 1197 to 934; on \dfpj{}, the corresponding averages decrease from 75.98\% to 64.77\%, from 58.12\% to 53.82\%, from 47.09\% to 44.55\%, and from 279 to 178. Although the drop is less severe than for usage retrieval, it remains substantial. This is consistent with the role of symbol constraints in preventing invalid references and hallucinated APIs during repair. Interestingly, symbol-level constraints can occasionally hurt individual project metrics.  For example, the full \method achieves a lower mutation score than \textit{-Sym} on Cli in \dfj{} and on A2A and ADK in \dfpj{}, and lower line or branch coverage on projects such as Cli, Codec, Gson, Jsoup, and ADK. Further investigation suggests that ambiguous symbol resolution is a plausible cause. For instance, ADK contains multiple classes with the same simple name across different packages and dependencies: \texttt{Schema} may refer to \texttt{com.google.genai.types.Schema}, \texttt{com.google.adk.tools.Annotations.Schema}, \texttt{com.google.cloud.bigquery.Schema}, or \texttt{org.apache.arrow.vector.types.pojo.Schema}. When the import context is incomplete, \method may deterministically prefer a project-local \texttt{com.google.adk.*} type over the intended external type due to package-prefix similarity. Thus, in projects with many homonymous APIs, symbolic resolution can occasionally select a wrong but plausible symbol. This limitation points to a future research direction: designing context-aware symbol resolution mechanisms that combine import history, call-site semantics, dependency usage patterns, and validation feedback to disambiguate symbols more reliably.

\textit{Protocol-level constraints} have the largest impact among the ablated components. The \textit{-Prot} variant achieves the lowest mutation score on both benchmarks (\SI{39.62}{\percent} and \SI{41.37}{\percent}) and the lowest line coverage on \dfj{} (\SI{77.77}{\percent}).
The largest drops occur on protocol-heavy projects such as Collections, JDatabind, A2A, JQuick, and JXml, where many CUTs involve stateful objects rather than simple method invocations. For example, JXml contains many stateful writer APIs whose deeper behavior is reachable only after satisfying specific protocol conditions. As discussed in Section~\ref{sec:intro}, a streaming writer must first establish the required name state by calling \texttt{setNextName(QName)} before invoking \texttt{writeStartObject()}.  Removing protocol-level constraints therefore makes it harder for the LLMs to follow the valid call orders, causing line coverage on JXml to drop from \SI{80.23}{\percent} to \SI{49.28}{\percent}, branch coverage from \SI{80.12}{\percent} to \SI{43.82}{\percent}, and mutation score from \SI{47.09}{\percent} to \SI{21.78}{\percent}. These results confirm that protocol-level constraints are especially important when the CUTs heavily depend on stateful objects.

Finally, \textit{iteration-level constraints} appear to have the weakest influence among the four ablated components, yet their contribution remains substantial. The \textit{-Iter} variant causes smaller drops than removing usage retrieval, symbol-level constraints, or protocol-level constraints, yet all metrics still decline substantially relative to the full method. On \dfj{}, line coverage decreases from 88.82\% to 80.81\%, branch coverage from 83.74\% to 74.51\%, mutation score from 52.00\% to 42.00\%, and DepLC from 1197 to 1006. On \dfpj{}, the corresponding averages decrease from 75.98\% to 65.94\%, from 58.12\% to 53.32\%, from 47.09\% to 45.97\%, and from 279 to 137. 
Unlike usage retrieval, symbol-level constraints, and protocol-level constraints, which are derived from source-code analysis, iteration-level constraints are learned from previous generation and repair experience. They operate across iterations by preserving useful experience, discouraging repeated failures, and reusing successful repair patterns. These results motivate a promising future direction: to incorporate richer memory mechanisms, such as MemoRAG~\cite{graphrag} or dual-memory designs~\cite{dualmemory}, to better organize past failures, successful repairs, dependency interactions, and reusable testing patterns.
\begin{center}
\fbox{\parbox{0.97\linewidth}{\textbf{Answer to RQ3:} 
All four components improve \method's effectiveness. Usage retrieval and protocol-level constraints have the largest impact, while symbol-level and iteration-level constraints provide additional gains by reducing hallucinated symbols and improving repair across iterations.
}}

\end{center}

\section{Threats to Validity}
\label{sec:threats}
\noindent\textbf{Construct Validity.}
A potential threat to construct validity is whether the selected metrics accurately reflect a test generation approach's effectiveness and efficiency. Standard metrics such as line coverage, branch coverage, and mutation score capture only part of test quality. For example, coverage may overestimate behavioral quality, while the mutation score can be influenced by the number and strength of assertions. To mitigate this threat, we use multiple complementary metrics, including line coverage, branch coverage, mutation score, project DepLC, number of iterations, token cost, and wall-clock time. Together, these metrics provide a more comprehensive view of both effectiveness and efficiency. In particular, we introduce the DepLC metric to measure how much real dependency code is exercised beyond the CUT, which is especially relevant for evaluating integration test generation.

\noindent\textbf{Internal Validity.}
Threats to internal validity mainly arise from the configurations of \method and PANTA, as well as the choice of the backbone LLM. First, different hyperparameter settings may affect the comparison between the two tools. To mitigate this threat, we use identical settings for both \method and PANTA whenever possible, including the same context size, temperature, maximum output length, LLM, tokenizer, and hardware environment. We also set the same generation budget for both tools, based on the experimental setting recommended by PANTA, to ensure they have sufficient opportunity to converge across all projects and demonstrate their full capabilities.   Second, the choice of LLM may influence the results. We select \texttt{Qwen3-Coder-30B-A3B-Instruct-FP8}, one of the latest and strongest small-scale open-source code models, considering both its coding capability and resource requirements. Larger LLMs may improve absolute performance, but benchmarking different LLMs is not the objective of this study, and newer LLMs may have been trained on the benchmark datasets, introducing further data leakage. Since both \method and PANTA use the same backbone model, stronger LLMs would likely benefit both approaches without altering the fairness of the comparison.

\noindent\textbf{Conclusion Validity.}
Threats to conclusion validity arise from the stochastic nature of LLM-based test generation. Even with the same input, an LLM may generate different tests across runs, which can affect the observed effectiveness and efficiency. To mitigate this threat, we report average results across projects and apply the Wilcoxon signed-rank test to assess whether the differences between \method and PANTA are statistically significant. We also report effect sizes using the Vargha--Delaney $\hat{A}_{12}$, so that the results reflect not only statistical significance but also the magnitude of the observed differences. Although repeated executions could further reduce the influence of randomness, repeating every experiment multiple times would substantially increase the cost of an already expensive evaluation. As reported in RQ2, the per-iteration cost is substantial for both \method and PANTA: on \dfj{}, each iteration consumes 61.42k tokens and 254.60 seconds for \method, compared with 50.17k tokens and 587.65 seconds for PANTA; on \dfpj{}, each iteration consumes 53.25k tokens and 147.22 seconds for \method, compared with 50.14k tokens and 222.58 seconds for PANTA. Each additional repetition would multiply LLM calls and repeat compilation, execution, coverage collection, mutation analysis, and repair. Given that our evaluation already covers many CUTs across diverse projects and shows consistent trends across benchmarks, we prioritize breadth across CUTs and projects over repeated executions of the same targets.

\noindent\textbf{External Validity.} Threats to external validity concern whether our results generalize to other Java projects. To mitigate this threat, we evaluate \method on \dfj{}, a widely used Java testing benchmark containing 14 projects from diverse domains. However, because \dfj{} was curated more than a decade ago, some of its code may have appeared in Qwen3-Coder's training corpus, raising a potential data leakage concern. To further assess generalizability and reduce this risk, we curate \dfpj{}, a new benchmark comprising Java projects created after the Qwen3-Coder training cutoff date. The consistent results of \method on both \dfj{} and \dfpj{} suggest that the observed improvements are not limited to a single benchmark or caused by benchmark memorization. Nevertheless, our evaluation remains limited to Java projects and a single backbone LLM, so future studies should extend to other languages, repositories, and models.

\section{Related Work}
\label{sec:related}
Software testing is a fundamental engineering task for ensuring software quality and mitigating release risks~\cite{llm4test}. We organize related work into two areas: automated unit test generation and automated integration test generation.

\subsection{Automated Unit Test Generation}
Early automated unit test generation approaches rely on search-based algorithms~\cite{evosuite2014,herlim2022citrus,lukasczyk2022pynguin,Pacheco2007Randoop:Java} and symbolic/concolic execution~\cite{chen2014test, garg2013feedback}. These approaches produce high-coverage tests, but their objectives are defined over the CUT alone.

Recent LLM-based test generation has received substantial attention~\cite{llm4test}. Existing approaches show that LLMs can generate complex inputs with domain-specific semantics~\cite{guzu2025large,huynh2025large}, construct effective test prefixes~\cite{pan2025aster,zhang2025citywalk,panta,candor}, and produce mock implementations for dependencies~\cite{gorla2025cubetesterai,roy2024static}. Below, we focus on the three approaches most closely related to \method in architecture or techniques.

\noindent\textit{CANDOR}~\cite{candor} is the first multi-agent, LLM-based test-generation framework for Java, decomposing the task into specialized agents (planner, tester, fixer). \method inherits this decomposition. However, CANDOR targets method-level programs without external dependencies (e.g., HumanEvalJava) and does not address the challenges of constructing and sequencing real dependency objects.

\noindent\textit{PANTA}~\cite{panta} advances LLM-based Java test generation to scenarios with dependencies, using path-guided prompting within an iterative generate-validate-fix loop. PANTA  shares the multi-agent, iterative design with \method. Despite these similarities, PANTA differs in three respects. First, PANTA's generation stage receives only path coverage and the CUT source code, without mining or injecting real dependency usage patterns from the project. Second, PANTA's repair stage relies purely on compilation and runtime error messages, without symbol-level, protocol-level, or iteration-level constraint enforcement. Third, PANTA does not explicitly target dependency coverage. Its tests commonly rely on Mockito in projects involving complex dependencies. Our mock usage analysis (Section~\ref{sec:rq1_results}) confirms that PANTA generates Mockito-based tests for 24\% of Deps4J CUTs.

\noindent\textit{Aster}~\cite{pan2025aster} generates test prefixes for Python using context retrieved from the codebase, refined through execution feedback. Aster shares the principle of context enrichment with \method, but differs in three ways. First, Aster targets Python whereas \method enforces strict Java type, import, and API protocol constraints. Second, Aster's context is general code snippets; \method's context is specifically structured as dependency usage patterns (instantiation chains, call sequences) mined via Joern-based program slicing. Third, Aster's refinement is feedback-driven but constraint-free, while \method enforces correctness through symbol indices, typestate models, and experience memory.

In summary, despite progress in LLM-based approaches, they commonly target unit-level testing, either operating at the method level without dependencies or relying on mocking to isolate the CUT. \method builds on CANDOR's multi-agent design and PANTA's path-guided prompting, but introduces two distinguishing mechanisms: context-enriched generation using mined dependency usage patterns, and constraint-enforced fixing across symbol, protocol, and iteration levels. None of the prior approaches are designed to exercise real dependency interactions for low-level integration test generation.

\subsection{Automated Integration Test Generation}
A separate line of work targets \emph{system-level} integration testing through REST APIs. EvoMaster~\cite{evomaster} uses evolutionary algorithms guided by OpenAPI specifications to generate HTTP requests maximizing service-level coverage. RESTler~\cite{restler} employs stateful fuzzing over OpenAPI descriptions. QuickREST~\cite{quickrest} leverages property-based testing over OpenAPI schemas. These tools operate at a fundamentally different level: they treat the system as a black box accessed through HTTP endpoints and use API specifications rather than source code. The challenges of object instantiation, import resolution, and intra-module API protocol compliance that motivate \method do not arise here.

Despite progress in both unit-level and system-level test generation, \emph{class-level} (low-level) integration test generation remains unaddressed. To the best of our knowledge, no prior automated approach explicitly targets generating tests that exercise a class together with its real, in-project dependencies at the source-code level. EvoSuite and Randoop incidentally cover dependency code, but their search objectives do not model dependency interactions. EvoMaster and similar tools target a different abstraction layer entirely. \method directly addresses this gap.
\section{Conclusion}
\label{sec:conclusion}
This paper presents \method, an LLM-based approach for generating low-level integration tests for Java. Unlike prior LLM-based approaches that target unit-level testing with mocked dependencies, \method directly exercises real project dependencies by combining context-enriched generation with constraint-enforced repair. It addresses two major sources of hallucination in dependency-aware test generation: \emph{not knowing}, mitigated by dependency-usage retrieval, and \emph{not following}, mitigated by symbol-, protocol-, and iteration-level constraints.

Our evaluation on \dfj{} and \dfpj{} shows that \method consistently improves test quality and dependency coverage over both the LLM-based SOTA baseline PANTA and the search-based SOTA baseline EvoSuite. On \dfj{}, \method improves average line coverage from 68.83\% to 88.82\% and from 61.52\% (EvoSuite) to 88.82\%, branch coverage from 58.84\% to 83.74\%, mutation score from 38.33\% to 52.00\%, and DepLC from 819 to 1197, while generating fewer tests than PANTA. On \dfpj{}, \method again improves line, branch, and dependency coverage, while maintaining comparable mutation scores. Although these gains require higher total generation cost in tokens and time, the cost remains practical, averaging 108.97 seconds and 26.59k tokens per method on \dfj{}, and 69.85 seconds and 25.46k tokens per method on \dfpj{}. An ablation study shows that all four major components of \method contribute positively, including usage retrieval, symbol-level constraints, protocol-level constraints, and iteration-level constraints.

Future directions include developing richer memory mechanisms to better organize past failures and successful repairs, improving context-aware symbol resolution for projects with many homonymous APIs, evaluating the readability and maintainability of generated tests, extending \method to other strongly typed languages, and investigating additional baselines such as larger LLMs and LLM-based coding agents.
\section*{Acknowledgement}
This work has emanated from research jointly funded by Taighde Éireann -- Research Ireland under Grant number~13/RC/2094\_2 and by Huawei Technologies Co., Ltd. Lionel Briand is also supported by the Natural Sciences and Engineering Research Council of Canada. For the purpose of Open Access, the authors have applied a CC BY public copyright licence to any Author Accepted Manuscript version arising from this submission.

\bibliography{references}

@book{Beck2022TestExample,
  title={Test driven development: By example},
  author={Beck, Kent},
  year={2002},
  publisher={Addison-Wesley Professional}
}

@article{evosuite2014,
author = {Fraser, Gordon and Arcuri, Andrea},
title = {A Large-Scale Evaluation of Automated Unit Test Generation Using EvoSuite},
year = {2014},
issue_date = {December 2014},
publisher = {Association for Computing Machinery},
address = {New York, NY, USA},
volume = {24},
number = {2},
issn = {1049-331X},
url = {https://doi.org/10.1145/2685612},
doi = {10.1145/2685612},
journal = {ACM Trans. Softw. Eng. Methodol.},
month = dec,
articleno = {8},
numpages = {42}
}

@misc{adk-java,
  title        = {Agent Development Kit for Java (ADK-Java)},
  author       = {{Google}},
  year         = {2025},
  howpublished = {\url{https://github.com/google/adk-java}},
  note         = {accessed March 2026}
}

@misc{fit-framework,
  title        = {FIT Framework: Model-Engineering Utilities for Java},
  author       = {{Huawei ModelEngine Group}},
  year         = {2025},
  howpublished = {\url{https://github.com/ModelEngine-Group/fit-framework}},
  note         = {accessed March 2026}
}

@misc{agentscope,
  title        = {AgentScope: A Flexible yet Robust Multi-Agent Platform (Java runtime)},
  author       = {{AgentScope Contributors}},
  year         = {2025},
  howpublished = {\url{https://github.com/agentscope-ai/agentscope-java}},
  note         = {accessed March 2026}
}

@misc{a2a-java,
  title        = {Agent-to-Agent (A2A) SDK for Java},
  author       = {{A2A Project Contributors}},
  year         = {2025},
  howpublished = {\url{https://github.com/a2aproject/a2a-java}},
  note         = {accessed March 2026}
}

@misc{jquick-curl,
  title        = {jQuick-Curl: An ANTLR-Based Curl Command Parser for Java},
  author       = {Pao, Haijiao},
  year         = {2025},
  howpublished = {\url{https://github.com/paohaijiao/jquick-curl}},
  note         = {accessed March 2026}
}

@inproceedings{lukasczyk2022pynguin,
  title={Pynguin: Automated unit test generation for python},
  author={Lukasczyk, Stephan and Fraser, Gordon},
  booktitle={Proceedings of the ACM/IEEE 44th International Conference on Software Engineering: Companion Proceedings},
  pages={168--172},
  year={2022}
}

@inproceedings{herlim2022citrus,
  title={Citrus: Automated unit testing tool for real-world c++ programs},
  author={Herlim, Robert Sebastian and Kim, Yunho and Kim, Moonzoo},
  booktitle={2022 IEEE Conference on Software Testing, Verification and Validation (ICST)},
  pages={400--410},
  year={2022},
  organization={IEEE Computer Society}
}

@article{llm4test,
  title={Large Language Models for Unit Test Generation: Achievements, Challenges, and Opportunities},
  author={Chu, Bei and Feng, Yang and Liu, Kui and Guo, Zhaoqiang and Zhang, Yichi and Shi, Hange and Nan, Zifan and Xu, Baowen},
  journal={arXiv preprint arXiv:2511.21382},
  year={2025}
}

@inproceedings{vllm,
  title={Efficient memory management for large language model serving with pagedattention},
  author={Kwon, Woosuk and Li, Zhuohan and Zhuang, Siyuan and Sheng, Ying and Zheng, Lianmin and Yu, Cody Hao and Gonzalez, Joseph and Zhang, Hao and Stoica, Ion},
  booktitle={Proceedings of the 29th symposium on operating systems principles},
  pages={611--626},
  year={2023}
}

@inproceedings{garg2013feedback,
  title={Feedback-directed unit test generation for C/C++ using concolic execution},
  author={Garg, Pranav and Ivan{\v{c}}i{\'c}, Franjo and Balakrishnan, Gogul and Maeda, Naoto and Gupta, Aarti},
  booktitle={2013 35th International Conference on Software Engineering (ICSE)},
  pages={132--141},
  year={2013},
  organization={IEEE}
}

@inproceedings{joern,
  author    = {Yamaguchi, Fabian and Golde, Nico and Arp, Daniel and Rieck, Konrad},
  title     = {Modeling and Discovering Vulnerabilities with Code Property Graphs},
  booktitle = {Proceedings of the 2014 IEEE Symposium on Security and Privacy (S\&P)},
  year      = {2014},
  pages     = {590--604},
  publisher = {IEEE},
  doi       = {10.1109/SP.2014.44}
}

@article{chen2014test,
  title={Test generation for embedded executables via concolic execution in a real environment},
  author={Chen, Ting and Zhang, Xiao-Song and Ji, Xiao-Li and Zhu, Cong and Bai, Yang and Wu, Yue},
  journal={IEEE Transactions on Reliability},
  volume={64},
  number={1},
  pages={284--296},
  year={2014},
  publisher={IEEE}
}

@misc{qwen3coder,
  title        = {{Qwen3-Coder}: Agentic Coding Models},
  author       = {{Qwen Team}},
  year         = {2025},
  howpublished = {\url{https://qwenlm.github.io/blog/qwen3-coder/}},
  note         = {Accessed March 2026}
}

@inproceedings{pit,
  author    = {Coles, Henry and Laurent, Thomas and Henard, Christopher and Papadakis, Mike and Ventresque, Anthony},
  title     = {{PIT}: A Practical Mutation Testing Tool for {Java} (Demo)},
  booktitle = {Proceedings of the 25th International Symposium on Software Testing and Analysis (ISSTA)},
  year      = {2016},
  pages     = {449--452},
  publisher = {ACM},
  doi       = {10.1145/2931037.2948707}
}

@article{panta,
  title={Llm test generation via iterative hybrid program analysis},
  author={Gu, Sijia and Nashid, Noor and Mesbah, Ali},
  journal={arXiv preprint arXiv:2503.13580},
  year={2025}
}

@article{sweabs,
  title={SWE-ABS: Adversarial Benchmark Strengthening Exposes Inflated Success Rates on Test-based Benchmark},
  author={Yu, Boxi and Cao, Yang and Zhang, Yuzhong and Lin, Liting and Xu, Junjielong and Zhong, Zhiqing and Xu, Qinghua and Wang, Guancheng and Cao, Jialun and Cheung, Shing-Chi and others},
  journal={arXiv preprint arXiv:2603.00520},
  year={2026}
}

@INPROCEEDINGS{mockornot,
  author={Spadini, Davide and Aniche, Maurício and Bruntink, Magiel and Bacchelli, Alberto},
  booktitle={2017 IEEE/ACM 14th International Conference on Mining Software Repositories (MSR)}, 
  title={To Mock or Not to Mock? An Empirical Study on Mocking Practices}, 
  year={2017},
  volume={},
  number={},
  pages={402-412},
  doi={10.1109/MSR.2017.61}}

@inproceedings{Fraser2011EvoSuite:Software,
  title={Evosuite: automatic test suite generation for object-oriented software},
  author={Fraser, Gordon and Arcuri, Andrea},
  booktitle={Proceedings of the 19th ACM SIGSOFT symposium and the 13th European conference on Foundations of software engineering},
  pages={416--419},
  year={2011}
}

@article{Hossain2024TOGLL:LLMs,
  title={Togll: Correct and strong test oracle generation with llms},
  author={Hossain, Soneya Binta and Dwyer, Matthew},
  journal={arXiv preprint arXiv:2405.03786},
  year={2024}
}

@inproceedings{Pacheco2007Randoop:Java,
  title={Randoop: feedback-directed random testing for Java},
  author={Pacheco, Carlos and Ernst, Michael D},
  booktitle={Companion to the 22nd ACM SIGPLAN conference on Object-oriented programming systems and applications companion},
  pages={815--816},
  year={2007}
}

@article{huynh2025large,
  title={Large language models for code generation: A comprehensive survey of challenges, techniques, evaluation, and applications},
  author={Huynh, Nam and Lin, Beiyu},
  journal={arXiv preprint arXiv:2503.01245},
  year={2025}
}

@article{guzu2025large,
  title={Large language models for c test case generation: A comparative analysis},
  author={Guzu, Alexandru and Nicolae, Georgian and Cucu, Horia and Burileanu, Corneliu},
  journal={Electronics},
  volume={14},
  number={11},
  pages={2284},
  year={2025},
  publisher={MDPI}
}

@inproceedings{pan2025aster,
  title={Aster: Natural and multi-language unit test generation with llms},
  author={Pan, Rangeet and Kim, Myeongsoo and Krishna, Rahul and Pavuluri, Raju and Sinha, Saurabh},
  booktitle={2025 IEEE/ACM 47th International Conference on Software Engineering: Software Engineering in Practice (ICSE-SEIP)},
  pages={413--424},
  year={2025},
  organization={IEEE}
}

@article{zhang2025citywalk,
  title={Citywalk: Enhancing llm-based c++ unit test generation via project-dependency awareness and language-specific knowledge},
  author={Zhang, Yuwei and Lu, Qingyuan and Liu, Kai and Dou, Wensheng and Zhu, Jiaxin and Qian, Li and Zhang, Chunxi and Lin, Zheng and Wei, Jun},
  journal={ACM Transactions on Software Engineering and Methodology},
  year={2025},
  publisher={ACM New York, NY}
}

@inproceedings{intention,
author = {Nan, Zifan and Guo, Zhaoqiang and Liu, Kui and Xia, Xin},
title = {Test Intention Guided LLM-Based Unit Test Generation},
year = {2025},
isbn = {9798331505691},
publisher = {IEEE Press},
url = {https://doi.org/10.1109/ICSE55347.2025.00243},
doi = {10.1109/ICSE55347.2025.00243},
booktitle = {Proceedings of the IEEE/ACM 47th International Conference on Software Engineering},
pages = {1026–1038},
numpages = {13},
location = {Ottawa, Ontario, Canada},
series = {ICSE '25}
}

@inproceedings{das2023comex,
  title={COMEX: A Tool for Generating Customized Source Code Representations},
  author={Das, Debeshee and Mathews, Noble Saji and Mathai, Alex and Tamilselvam, Srikanth and Sedamaki, Kranthi and Chimalakonda, Sridhar and Kumar, Atul},
  booktitle={2023 38th IEEE/ACM International Conference on Automated Software Engineering (ASE)},
  pages={2054--2057},
  year={2023},
  organization={IEEE}
}

@article{dualmemory,
  title={Experepair: Dual-memory enhanced llm-based repository-level program repair},
  author={Mu, Fangwen and Wang, Junjie and Shi, Lin and Wang, Song and Li, Shoubin and Wang, Qing},
  journal={arXiv preprint arXiv:2506.10484},
  year={2025}
}

@inproceedings{graphrag,
  title={Memorag: Boosting long context processing with global memory-enhanced retrieval augmentation},
  author={Qian, Hongjin and Liu, Zheng and Zhang, Peitian and Mao, Kelong and Lian, Defu and Dou, Zhicheng and Huang, Tiejun},
  booktitle={Proceedings of the ACM on Web Conference 2025},
  pages={2366--2377},
  year={2025}
}

@article{a12,
  title={A critique and improvement of the CL common language effect size statistics of McGraw and Wong},
  author={Vargha, Andr{\'a}s and Delaney, Harold D},
  journal={Journal of Educational and Behavioral Statistics},
  volume={25},
  number={2},
  pages={101--132},
  year={2000},
  publisher={Sage Publications Sage CA: Los Angeles, CA}
}

@inproceedings{gorla2025cubetesterai,
  title={CubeTesterAI: Automated JUnit Test Generation Using the LLaMA Model},
  author={Gorla, Daniele and Kumar, Shivam and Lorenzini, Pietro Nicolaus Roselli and Alipourfaz, Alireza},
  booktitle={2025 IEEE Conference on Software Testing, Verification and Validation (ICST)},
  pages={565--576},
  year={2025},
  organization={IEEE}
}

@inproceedings{roy2024static,
  title={Static program analysis guided LLM based unit test generation},
  author={Roy Chowdhury, Sujoy and Sridhara, Giriprasad and Raghavan, AK and Bose, Joy and Mazumdar, Sourav and Singh, Hamender and Sugumaran, Srinivasan Bajji and Britto, Ricardo},
  booktitle={Proceedings of the 8th International Conference on Data Science and Management of Data (12th ACM IKDD CODS and 30th COMAD)},
  pages={279--283},
  year={2024}
}

@misc{2025LangchainWebsite,
title = "Langchain Official Website",
year = "Accessed: 2025",
url = {https://www.langchain.com/}
}

@inproceedings{defects4j,
author = {Just, Ren\'{e} and Jalali, Darioush and Ernst, Michael D.},
title = {Defects4J: a database of existing faults to enable controlled testing studies for Java programs},
year = {2014},
isbn = {9781450326452},
publisher = {Association for Computing Machinery},
address = {New York, NY, USA},
url = {https://doi.org/10.1145/2610384.2628055},
doi = {10.1145/2610384.2628055},
booktitle = {Proceedings of the 2014 International Symposium on Software Testing and Analysis},
pages = {437–440},
numpages = {4},
location = {San Jose, CA, USA},
series = {ISSTA 2014}
}

@inproceedings{Arcuri2011AEngineering,
  title={A practical guide for using statistical tests to assess randomized algorithms in software engineering},
  author={Arcuri, Andrea and Briand, Lionel},
  booktitle={Proceedings of the 33rd international conference on software engineering},
  pages={1--10},
  year={2011}
}

@article{candor,
  title={Hallucination to Consensus: Multi-Agent LLMs for End-to-End JUnit Test Generation},
  author={Xu, Qinghua and Wang, Guancheng and Briand, Lionel and Liu, Kui},
  journal={ACM Transactions on Software Engineering and Methodology},
  year={2026},
  publisher={ACM New York, NY}
}

@article{mutgen,
  title={Mutation-Guided Unit Test Generation with a Large Language Model},
  author={Wang, Guancheng and Xu, Qinghua and Briand, Lionel C and Liu, Kui},
  journal={arXiv preprint arXiv:2506.02954},
  year={2025}
}

@misc{mockito,
  author       = {Szczepan Faber and Brice Dutheil and Rafael Winterhalter and Tim van der Lippe and others},
  title        = {Mockito},
  howpublished = {\url{https://mockito.org/}},
  year         = {n.d.},
  note         = {Mocking framework for unit tests in Java}
}

@misc{powermock,
  author       = {Johan Haleby and Jan Kronquist and others},
  title        = {PowerMock},
  howpublished = {\url{https://github.com/powermock/powermock}},
  year         = {n.d.},
  note         = {Extension to Mockito/EasyMock for mocking static/final/private methods, etc.}
}

@INPROCEEDINGS{evomaster,
  author={Arcuri, Andrea},
  booktitle={2018 IEEE 11th International Conference on Software Testing, Verification and Validation (ICST)}, 
  title={EvoMaster: Evolutionary Multi-context Automated System Test Generation}, 
  year={2018},
  volume={},
  number={},
  pages={394-397},
  doi={10.1109/ICST.2018.00046}}

@INPROCEEDINGS{restler,
  author={Atlidakis, Vaggelis and Godefroid, Patrice and Polishchuk, Marina},
  booktitle={2019 IEEE/ACM 41st International Conference on Software Engineering (ICSE)}, 
  title={RESTler: Stateful REST API Fuzzing}, 
  year={2019},
  volume={},
  number={},
  pages={748-758},
  doi={10.1109/ICSE.2019.00083}}

@inproceedings{dynamosa,
  title={Extending AVM with genetic algorithm and dynamic fitness landscape analysis for unit test case generation},
  author={Panichella, Annibale and Kifetew, Fitsum Meshesha and Tonella, Paolo},
  booktitle={Proceedings of the 2015 IEEE/ACM 10th International Workshop on Search-Based Software Testing (SBST)},
  pages={9--12},
  year={2015},
}

@article{quickrest,
  title={QuickREST: Property-based Test Generation of OpenAPI-Described RESTful APIs},
  author={Stefan Karlsson and Adnan Causevic and Daniel Sundmark},
  journal={2020 IEEE 13th International Conference on Software Testing, Validation and Verification (ICST)},
  year={2019},
  pages={131-141},
  url={https://api.semanticscholar.org/CorpusID:209439495}
}

@String{Computing = "Computing" }

@String{Computer = "{IEEE} Computer" }
\bibliographystyle{ACM-Reference-Format}


\end{document}